\documentclass[prd,twocolumn,showpacs,preprintnumbers,nofootinbib,prd,superscriptaddress,groupedaddress,10pt]{revtex4-1}

\usepackage{amsmath,amssymb}
\usepackage{amsfonts}
\usepackage{bm}
\usepackage{graphics}
\usepackage{float}
\usepackage{amsmath}
\usepackage{amssymb}
\usepackage[normalem]{ulem} %for \sout
\usepackage[mathscr]{euscript}
\usepackage{acronym}
\usepackage{float}
\usepackage[caption=false]{subfig}
\usepackage{lipsum}
\usepackage{mathtools}
\usepackage{multirow}
\usepackage{graphicx}
\usepackage[usenames,dvipsnames]{xcolor}
\usepackage[colorlinks, pdfborder={0 0 0}]{hyperref} % should always be loaded last
\usepackage[nameinlink,capitalise]{cleveref}

\definecolor{LinkColor}{rgb}{0.75, 0, 0}
\definecolor{CiteColor}{rgb}{0, 0.5, 0.5}
\definecolor{UrlColor}{rgb}{0, 0, 0.75}
\hypersetup{linkcolor=LinkColor}
\hypersetup{citecolor=CiteColor}
\hypersetup{urlcolor=UrlColor}
\graphicspath{{fig}}
\usepackage{booktabs}

\newcommand{\Msun}{\ensuremath{M_\odot}}

\begin{document}
 
\title{Ringdown overtones, black hole spectroscopy, and no-hair theorem tests}
%%%%
\author{
Swetha Bhagwat$^1$,
Xisco Jim\'enez Forteza$^{1,2,3}$,
Paolo Pani$^{1}$,
Valeria Ferrari$^1$}

\affiliation{$^1$ Dipartimento di Fisica, ``Sapienza'' Universit\`a di Roma \& Sezione INFN Roma1, Piazzale Aldo Moro 
5, 00185, Roma, Italy}
\affiliation{$^2$ Max Planck Institute for Gravitational Physics (Albert Einstein Institute), Callinstra{\ss}e 38, 30167 Hannover, Germany}
\affiliation{$^3$ INFN Sez. di Napoli, Compl. Univ.  Monte S. Angelo Ed. G, Via Cinthia, I-80126 Napoli (Italy)}

\begin{abstract}
Validating the black-hole no-hair theorem with gravitational-wave observations of compact binary coalescences provides a compelling argument that the remnant object is indeed a black hole as described by the general theory of relativity. This requires performing a spectroscopic analysis of the post-merger signal and resolving the frequencies of either different angular modes or overtones (of the same angular mode). For a nearly-equal mass binary black-hole system, only the dominant angular mode ($l=m=2$) is sufficiently excited and the overtones are instrumental to perform this test. Here we investigate the robustness of modelling the post-merger signal of a binary black hole coalescence as a superposition of overtones. Further, we study the bias expected in the recovered frequencies as a function of the start time of a spectroscopic analysis and provide a computationally cheap procedure to choose it based on the interplay between the expected statistical error due to the detector noise and the systematic errors due to waveform modelling. Moreover, since the overtone frequencies are closely spaced, we find that resolving the overtones is particularly challenging and requires a loud ringdown signal. Rayleigh's resolvability criterion suggests that --~in an optimistic scenario~-- a ringdown signal-to-noise ratio larger than $\sim 30$ (achievable possibly with LIGO at design sensitivity and routinely with future interferometers such as Einstein Telescope, Cosmic Explorer, and LISA) is necessary to resolve the overtone frequencies. We then conclude by discussing some conceptual issues associated with black-hole spectroscopy with overtones.
\end{abstract}
\maketitle

\section{Introduction}
\label{sec:intro}
At a sufficiently late time after the merger, the evolution of a binary black hole~(BBH) spacetime can be described as a perturbation on the spacetime of the remnant black hole~(BH). The gravitational-wave~(GW) signal produced during these final stages is called the `ringdown'~(RD)\footnote{Throughout this paper RD is assumed to be the part of the BBH waveform describable by linear perturbation theory of the remnant BH. Note that in literature, RD is also sometimes used to refer to the full post-merger signal which contrasts with the usage of the word we choose here.}. It comprises of a superposition of damped sinusoids with frequencies and damping times corresponding to the  quasi-normal-mode~(QNM) spectrum~\cite{Chandrasekhar,10.2307/2397876,Kokkotas:1999bd,Berti:2009kk,Vishveshwara:1970zz,Teukolsky1,Teukolsky2,Teukolsky3} of the remnant BH.
\begin{align}
\label{eq:rdmodel}
    h = \Sigma_{lmn} {\cal A}_{lmn} e^{- \iota \omega_{lmn} t} e^{- t/ \tau_{lmn}}\, \mathcal{Y}_{lm}\,,
\end{align}
where the QNMs are indexed by a set of three integers: ($l,m$) describe the angular dependence of the modes and are decomposed on the spin-weighted $s=2$ spheroidal harmonics $\mathcal{Y}_{lm}$, whereas $n$ is the overtone index of a given angular mode, with $n=0$ being the fundamental mode. For a given QNM, the complex amplitude ${\cal A}_{lmn}=A_{lmn}e^{i\phi_{lmn}}$ depends on the perturbation conditions set up during the inspiral-plunge-merger phase of the BBH evolution. The QNM frequencies $\omega_{lmn}$ and the damping times $\tau_{lmn}$ are characteristic of the final BH and are parametrized uniquely by its mass~($M_{f}$) and spin~($a_{f}$).

The BH uniqueness~\cite{Robinson:2004zz, Mazur:1982db,Gibbons:2002av,Israel:1967wq} and the no-hair theorem~\cite{Figueras:2009ci,Carter:1971zc,Hawking:1973uf} state that a BH spacetime in general relativity~(GR) is uniquely defined by at most three parameters: the mass, the angular momentum and the electric charge. The latter is thought to be negligible for astrophysical BHs. All aspects in a BH spacetime --~including its perturbative dynamics and therefore the QNM spectrum~-- is thus completely parametrizable by its mass and spin. Consequently, the countably infinite number of QNMs of a Kerr BH are uniquely related to each other. This presents a lucrative opportunity to perform an observational validation --~in the form of multiple null-hypothesis tests~-- of the no-hair theorem with the BBH RD signals~\cite{Chirenti:2017mwe,TGR,Li:2011cg}.  

The dominant QNM excitation in a BBH merger is the $l=m=2, n=0$ QNM and is the easiest to detect in the GW data. To perform a no-hair theorem test one needs to at least obtain an independent measurement of three QNM parameters.  In particular, one can measure the fundamental $l=m=2$ mode along with either
%%%
\begin{itemize}
    \item  another angular mode, namely the $l=m=3$ or the $l=2,\, m=1$ mode, both with $n=0$; or
    \item  the first overtone ($n=1$) of the $l=m=2$ mode.
\end{itemize}
Traditionally, BH spectroscopy and the no-hair theorem tests have been performed using different angular modes in the RD~\cite{TGR,greg1,greg2,Baibhav,Bhagwat:2016ntk,2019arXiv191013203B,Gossan:2011ha}.

\subsubsection*{Role of the overtones for BH spectroscopy and no-hair theorem tests}

Although the role of overtones in RD modelling has been studied for a long time \cite{London:2014cma,Taracchini:2013rva}, they have been considered for performing the no-hair theorem test only recently~\cite{Isi:2019aib,Matt}. Overtones have become particularly important because a substantial fraction of the BBHs detected by the LIGO-VIRGO detectors comprise of near-equal mass systems~\cite{LIGOScientific:2018mvr}. This poses a predicament for using angular modes to perform the no-hair theorem tests as these systems do not sufficiently excite angular modes other than the dominant one (e.g., the $l=m=3$ or $l=2, m=1$ modes are suppressed)~\cite{Berti:2007zu,Brito:2018rfr}. For these systems, performing a no-hair theorem test using the overtones of $l=m=2$ mode seems promising. 

The LIGO-Virgo Collaboration performed a GR consistency test by comparing the full inspiral-merger-RD waveform to the RD of the remnant BH modelled by a single dominant mode for GW150914 event~\cite{TGR}.
Recent work like Refs.~\cite{Isi:2019aib,Matt} tried to quantify the contribution of additional overtones to the RD signal of GW150914 event. Adding at least one overtone is necessary to obtain an independent estimate of the remnant BH's mass and spin from the RD of GW150914 system. It was found that these estimates were consistent (at the $\sim 20\%$ level) with the values predicted by numerical relativity~(NR) simulations corresponding to the estimated BBH parameters of a full inspiral-merger-RD  waveform~\cite{Matt,Isi:2019aib} (see
also~\cite{Berti:2007zu,Baibhav:2017jhs}).

However, going beyond consistency tests demands a robust \emph{BH spectroscopy}, i.e., measuring several QNMs from a single BBH RD. An observational validation of the no-hair theorem requires that the QNM frequencies and the damping times must be estimated by performing a Bayesian parameter estimation~(PE) on all the RD model parameters\footnote{Henceforth, for ease of notation, we drop the ($l,m$) indices as we only study the overtones of the $l=m=2$ angular mode. For instance, ${A}_n\equiv A_{22n}$ refers to the amplitude of the $l=m=2$ mode with overtone index $n$, where $n=0$ is the fundamental mode.} (i.e., $\{ {A}_{n}, \phi_{n}, \omega_{n}, \tau_{n}\}$) for a BBH RD event. It is required that: a) the obtained estimates of the frequencies and damping times are consistent with the predicted GR-QNM spectrum; and b) the QNM frequency spectrum can be resolved, i.e., that the separation of the two estimated frequencies are larger than the measurement errors, for the two frequencies to be distinguishable.

We consider the RD waveform produced by NR simulations as a gold standard and compare them with analytical RD models. Accurate modelling of the postmerger\footnote{Note that although the peak of the waveform amplitude is often loosely referred to as the time of the merger, this time does not necessarily correspond to the merger of the horizon of the two progenitor-BHs. Nonetheless, in this paper we refer to the time at the peak of the waveform as the time of the merger.} allows for a better estimation of the mass and the spin of the remnant BH (thus allowing for more accurate inspiral-merger-RD consistency tests~\cite{TGR,Isi:2019aib,Matt}). 

Details of the fitting procedure and some discussion related to this fitting are presented in Sec.~\ref{sec:fit}. Including up to $n = 7$ overtones in the RD waveform is necessary to model the post-merger accurately. However an 8-tone\footnote{We define a $p$-tone RD model as that including $n=0,1,2,...,p-1$ overtones. Thus, an 8-tone model includes overtones from $n=0$ to $n=7$, whereas a $2$-tone model includes only the fundamental mode and the $n=1$ overtone.} RD model has $4 \times 8 = 32$ independent model parameters which makes performing a Bayesian PE infeasible. The simplest Bayesian PE using overtones can be performed with a \emph{minimal} RD model comprising of a fundamental mode and its first overtone. In this paper, we study this model and refer to it as the \emph{$2$-tone RD model} henceforth.

The $2$-tone model is insufficient to describe the entire post-merger accurately~\cite{Isi:2019aib,Matt,Baibhav:2017jhs} and we discuss it in Sec.~\ref{sec:time}. It is important to start the analysis from an appropriate time, $t_0$, after the merger such that the recovered QNMs do not have a bias arising from the modelling inaccuracies. In Sects.~\ref{sec:time} and~\ref{sec:prescription} we address this issue and provide a computationally cheap recipe to choose the start time for a spectroscopic analysis of a BBH RD using overtones\footnote{Our prescription can be easily extended to BH spectroscopy with different angular modes.}. 

Performing the no-hair theorem test with overtones is challenging, both conceptually and from an implementation point of view. Although the addition of overtones improves the accuracy of RD waveform models, whether they can be used for performing a direct no-hair theorem test must be investigated carefully, particularly because the overtone frequencies are closely spaced and a high signal-to-noise ratio~(SNR) is required to resolve them. In Sec.~\ref{sec:resolvability} we provide an estimate of the minimum SNR required to resolve overtones of a BBH system based on Rayleigh's resolvability criterion (the latter providing an optimistic \emph{lower} bound for the minimum SNR~\cite{Berti:2007zu}) required.  
Finally, in Sect.~\ref{sec:discussion} we conclude by discussing some conceptual issues that must be considered while performing BH spectroscopy with overtones.

 %%%%%%%%%%%%%%%%%%%%%%%%%%%%%%%%%%%%%%%%%%%%%
\section{Fitting the RD with overtones}
\label{sec:fit}
 %%%%%%%%%%%%%%%%%%%%%%%%%%%%%%%%%%%%%%%%%%%%%

\begin{figure}[h!]
 \subfloat{\includegraphics[width=0.45\textwidth]{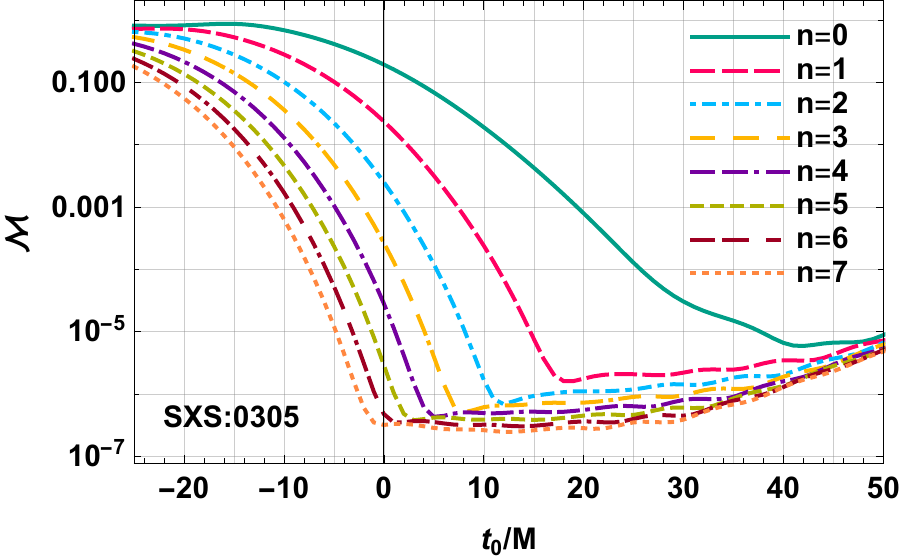}} \\
 \subfloat{\includegraphics[width=0.45\textwidth]{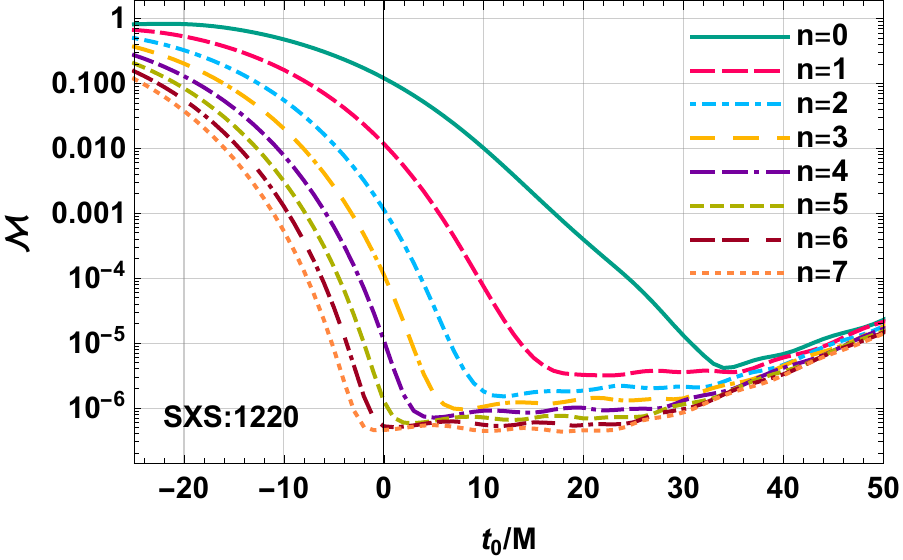}}  
 \caption{Mismatch ${\cal M}$ of multi-tone RD model relative to NR waveforms (SXS:0305 and SXS:1220) as a function of the start time $t_0$~\cite{Matt, Isi:2019aib}. Different curves correspond to the number of overtones included.
 Note that the curves for the two different BBH systems are qualitatively similar.
 } 
 \label{fig:matchplot}
 \end{figure}
 
Estimating the amplitudes and the phases of QNMs in a BBH RD from the source frame dynamics is a highly nontrivial problem. In practice, a fit must be performed to calibrate the amplitudes and the phases of a multi-tone RD model against an NR-RD. To validate our fitting procedure and to check the reproducibility of the best fit values obtained for the amplitudes and the phases of an 8-tone RD model, we first compare our fitting results with those in Ref.~\cite{Matt}. This is particularly important because the overtones are short-lived and therefore the fits could be unstable and/or irreproducible.

We consider 2 nearly non-spinning BBH NR simulations from the SXS catalogue~\cite{Boyle:2019kee} as our representative systems: (i)~SXS:0305 corresponding to mass ratio $q=1.2$, and (ii)~SXS:1220 corresponding to $q=4$. The spins of the remnant BHs are $a_{f}^{\rm SXS:0305}=0.69$ and $a_{f}^{\rm SXS:1220}=0.47$, respectively. The NR simulations scale with the total mass of the system and in this section, we normalize it to unity. We fix the frequency spectrum of the damped sinusoids in Eq.~\eqref{eq:rdmodel} to the GR-QNM spectrum of the remnant BH for various overtones of the $l=m=2$ angular mode. We then use a complex linear least-squared fit based on $\chi^2$ minimisation to obtain the best fit value of the amplitudes and phases for our RD models.  $\chi^2$ is defined as 
\begin{equation}
\chi^2=\sum_{i} |\bar{h}(\vec{\lambda})_i-h_i|^2,
\end{equation}
where $\bar{h}(\vec{\lambda})_i$ is the model parameterized by $\vec{\lambda}=\left\lbrace A_{lmn}; \omega_{lmn},\tau_{lmn}, t_0 \right\rbrace$ and is evaluated at each time step $t=t_i$. 

Throughout this study, while we fit the RD at different start time $t=t_0$, we quote the amplitude extrapolated back in time at $t=0$ for the ease of comparison. This is done by rescaling the amplitude as $A_{lmn}\to$ ${A}_{lmn}e^{(\iota \omega_{lmn} + 1/\tau_{lmn})t_0}$. Further, for the sake of simplicity, we assume that all overtones are excited simultaneously. We emphasise this is a strong assumption in Sec.~\ref{sec:discussion}.

\begin{table}
\begin{tabular}{|c|c|c|c|c|}
\hline
 & \multicolumn{2}{c|}{SXS:0305}& \multicolumn{2}{c|}{SXS:1220} \\
 \hline
\textbf{n}& $A_{n}$ & $\phi_{n}$ & $A_{n}$ & $\phi_{n}$\\ \hline
0 & 0.978518 & -2.11289 & 0.633972 & -0.475193 \\ \hline
1 & 4.29435 & 1.38519 & 2.01749 & -1.57366\\ \hline
2 & 11.6503 & 0.128732 & 4.25772 & 0.725335\\ \hline
3 & 23.6475 & 2.24624 & 7.47324 & -2.35637,\\ \hline
4 & 34.0133 & -0.0202084 & 10.2204 & -1.83581\\ \hline
5 & 30.4153 & 0.963395 & 9.47404 & -1.35298\\ \hline
6 & 14.8093 & -0.797055 & 5.05658 & -0.330465\\ \hline
7 & 3.03776 & -2.64276 & 1.15199 & -2.27162 \\
\hline
\end{tabular}
\caption{The best-fit amplitudes and phases of an 8-tone RD model. We perform the fits at the peak of the BBH waveform $t_0=0$ for SXS:0305 and SXS:1220 and provide the values of the best fit amplitudes and phases. We note that for SXS:0305 these values are comparable to that presented in Ref.~\cite{Matt}.}
\label{tab:Fit-SXS0305-7overtones}
\end{table}

With these assumptions, we perform the complex linear RD fits and in Table~\ref{tab:Fit-SXS0305-7overtones} we present the best-fit values obtained for the amplitudes and the phases of QNMs. We confirm that the best fit amplitude values corresponding to SXS:0305 are consistent with that presented in Table~1 of Ref.~\cite{Matt}.

From Table~\ref{tab:Fit-SXS0305-7overtones}, we note that the largest excitation amplitude in both the NR simulations considered here corresponds to the $n=4$ overtone\footnote{We observe the magnitude of the best fit amplitude first increases, reaches the maximum at $n=4$ and then decreases. Here is a speculative explanation for this phenomenon. The first overtones to be excited are the high-$n$ ones, which are excited
by sources relatively far from the horizon and therefore their
excitation factor is small. On the other hand, low-$n$ overtones are
excited by sources near the horizon, so part of the ringdown energy is
absorbed and never reaches infinity. The largest excitation occurs for intermediate-$n$ overtones, which are not too close nor too distant from the horizon.} The amplitudes of the overtones higher than $n=4$ decrease monotonically. Although not conclusive, this hints towards the possibility that only a finite number of overtones are excited significantly in a BBH RD. 

To quantify the accuracy of our fits, we use the mismatch defined as
\begin{equation}
    \mathcal{M} = 1 - \mathcal{C}\,,\label{mismatch}
\end{equation}
where $ \mathcal{C}$ is the `averaged match' defined as
\begin{align}
    \mathcal{C} = \frac{\int_{t_0}^{t_{\rm end}} h_{\rm NR}(t) h_{M}(t) dt }{\sqrt{\int_{t_0}^{t_{\rm end}} h_{\rm NR}(t) h_{\rm NR}(t) dt}\times \sqrt{\int_{t_0}^{t_{\rm end}} h_{M}(t) h_{M}(t) dt}}. 
\end{align}
Here $h_{\rm NR}$ represents the NR-RD while $h_{M}$ is a $p$-tone RD model (obtained by truncating Eq.~\eqref{eq:rdmodel} at $n=p-1$), whereas $t_0$ is the time after the peak of the GW waveform at which we begin our fits and \textcolor{BrickRed}{$t_{end}=90M$}. 

In Fig.~\ref{fig:matchplot} we show the behaviour of $\mathcal{M}$ as a function of $t_0$ for the $p$-tone RD models. The top panel shows an overall agreement with Fig.~1 in \cite{Matt}, validating our fitting procedure and confirming the reproducibility of the best-fit amplitude and phase values for SXS:0305. Further, we observe that the behaviours of the mismatch for SXS:0305 and SXS:1220 are qualitatively similar. We find that in both cases, the entire post-merger is reproduced accurately (${\cal M}\sim 10^{-6}$) by an 8-tone RD model\footnote{At late times ($\sim t\gtrsim60M$) the fit results are highly sensitive to the numerical noise in the NR simulations. We observe a rising trend in the mismatch ${\cal M}$ at late times in Fig. \ref{fig:matchplot} that seems to be inconsistent with Fig 1 of \cite{Matt}. However, we want to highlight that this regime is noise dominated and thus the fitting is unstable. Furthermore, we find that by setting $t_{end}=60M$ in the mismatch computation allows us to recover the same flatter profile at late times similar to one obtained in \cite{Matt}.}. 

\begin{figure}
\subfloat{\includegraphics[width=0.45\textwidth]{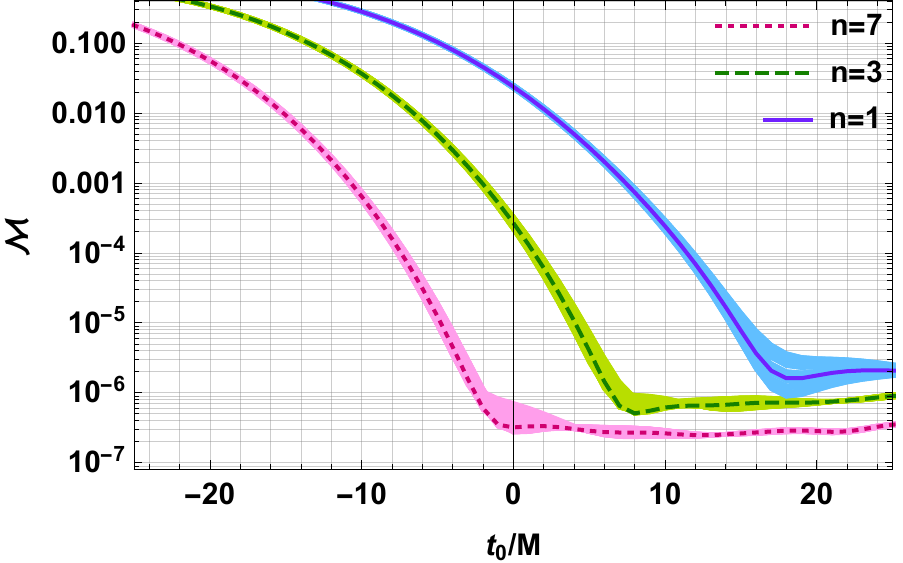}} 
\caption{Mismatch of RD models with non-GR-QNM spectrum for SXS:0305. We computed the mismatch (similar to Fig.~\ref{fig:matchplot}) of the best-fit waveform corresponding to a $2$-tone, $4$-tone and $8$-tone RD models (overtones up to $n=1,3,7$ respectively, see legend) with NR-RD. The shaded bands show the mismatch as a function of time for $100$ realisations of the RD-models with non-GR QNM spectra. The frequencies of the overtones ($n>0$) are randomly drawn within $2\%$ the GR-QNM frequencies. The solid curves correspond to the RD models with GR-QNM spectrum. Note that all the three solid curves lie within their respective shaded bands demonstrating that some non-GR-QNM spectra produce a similar (or better) mismatch.
}
\label{fig:matchplotrand}
\end{figure}

However, we remark that although with an 8-tone RD model the post-merger signal is accurately replicated~\cite{Matt,Isi:2019aib}, to interpret them as actual BH QNMs one needs to verify that the choice of the spectrum is uniquely specified by GR and that a different choice of the frequency spectrum cannot reproduce the post-merger signal with a similar accuracy. This is crucial to avoid the risk of misinterpreting the fit, especially since the number of parameters of the model is large. Therefore, in Fig.~\ref{fig:matchplotrand} we investigate if frequencies inconsistent with the GR-QNM spectrum can reproduce SXS:0305 NR-RD accurately. We fix the frequency and damping time of the fundamental mode in an $p$-tone RD model because the frequency of the fundamental mode is required to reproduce the late stages of RD. However, we allow the frequencies of all the remaining $p-1$ overtones to randomly vary within $\pm 2 \%$ of their GR QNM values\footnote{Note that the difference in frequencies  between the fundamental mode and the $n=1$ overtone for a SXS:0305-like system is $\sim 2.2 \%$ of the $l=m=2, n=0$ QNM frequency.}. We fit for the amplitudes and the phases of the overtones to the NR-RD. 

We draw 100 realisations of non-GR-QNM frequency spectra and repeat this analysis on each of them. The shaded bands of Fig.~\ref{fig:matchplotrand} show the mismatch of 100 realisations of frequency spectra for $2$-tone, $4$-tone and $8$-tone RD models. The solids lines in the figure correspond the $p$-tone RD models (with overtones up to $n=1,3,7$, respectively) using the GR-QNM frequency spectrum. 

We find that the mismatch obtained with a GR-QNM spectrum lies within the shaded bands, showing that non-GR-QNM spectra can reproduce an NR-RD to a similar accuracy as GR-QNMs. In fact, for each model there exists a few realisations of the non-GR-QNM spectrum that reproduce the NR-RD to a slightly better accuracy than the GR-QNM spectrum. This casts some doubts on the interpretation of the damped sinusoid as BH QNMs\footnote{The fact that some effective-one-body waveform models use pseudo-modes (i.e., modes which are not QNMs) to reproduce the early RD~\cite{Taracchini:2013rva} provides further hint that these damped sinusoids may just be basis functions producing reliable fits.}. 

Moreover, we also find that the best fit amplitudes $A_n$ (particularly, for the higher $n$ overtones) are extremely sensitive to small changes in the overtone frequencies. This is shown in Appendix~\ref{app:correlation} (cf. Fig.~\ref{fig:amps_variation}).

An $8$-tone RD model is impractical for a Bayesian PE because one needs to estimate $8 \times 4=32$ parameters from a few cycles of RD. Therefore, in the rest of this paper, we concentrate on a more feasible $2$-tone RD model --~including only the fundamental mode and the $n=1$ overtone~-- which is the minimal complexity of RD model required to perform a no-hair theorem test. It must be emphasised that a $2$-tone RD model is insufficient to capture the morphology of NR-RD close to the peak of the waveform~\cite{Baibhav:2017jhs} and one is forced to choose an appropriate RD start time $t_0$.  We elaborate on this in Sect.~\ref{sec:time}.

\begin{figure}
\subfloat{\includegraphics[width=0.45\textwidth]{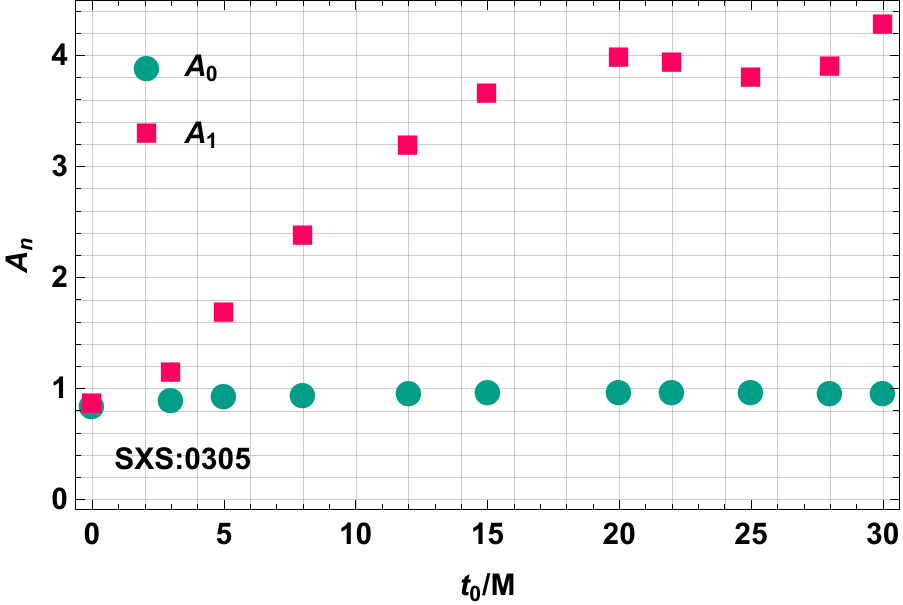}}\\
\subfloat{\includegraphics[width=0.45\textwidth]{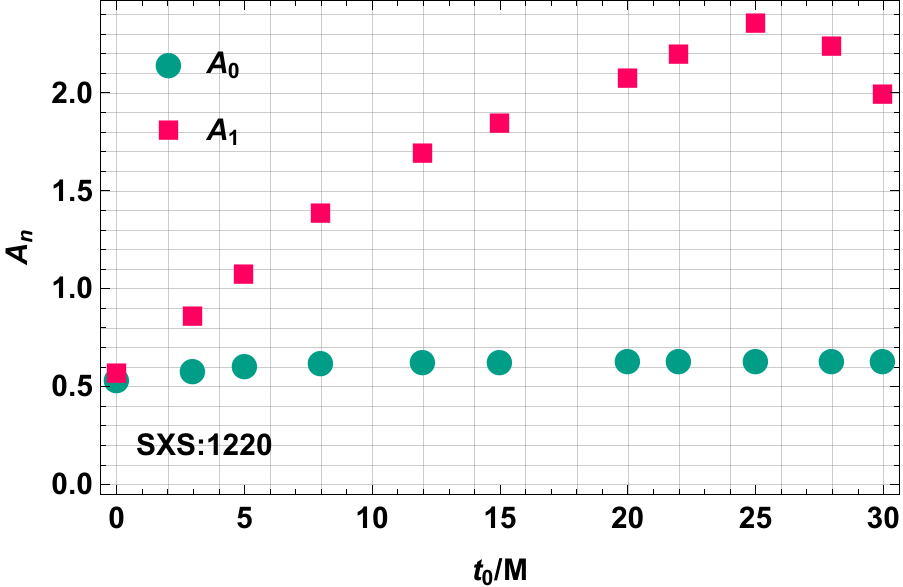}} 
\caption{Best fit amplitude for a $2$-tone RD model as a function of $t_{0}$ for SXS:0305 (top panel) and SXS:1220 (bottom panel). The best fit value of amplitudes ${A}_{0}$ (green dotted markers) and ${A}_{1}$ (red squared markers) for the $2$-tone RD-model are presented as a function of start time of the fit $t_0$. For ease of comparison, we report the values of amplitudes at the merger time obtained by extrapolating the best fit amplitude at $t_{0}$ to $t=0$. Ideally, when the best fit amplitudes are robust to the start time of fits, one expects the markers to line up at a constant value. We see that while the best-fit amplitude $A_{0}$ is robust to the choice of $t_0$, $A_{1}$ increases, indicating an instability towards the choice of start time. 
}
\label{fig:ampstab}
\end{figure}

Finally, we investigate the robustness of the fits for a $2$-tone RD model by studying the behaviour of the best-fit amplitudes ${A}_{0,1}$ with the start time $t_0$ (specifically, $t_0/M \in [0,30]$) for SXS:0305 and SXS:1220. In Fig.~\ref{fig:ampstab}, the markers correspond to the values for the amplitudes ${A}_{0,1}$ obtained by fitting a $2$-tone RD model using the GR-QNM spectrum. For both the systems, the best-fit amplitude ${A}_0$ (shifted back to the merger time) does not vary considerably with the start time of the fit. This indicates that the dominant $n=0$ mode is robust to variations of $t_0$. However, the best fit value for ${A}_{1}$ increases significantly for larger values of $t_0$, alluding to an instability towards the choice of start time of the fits. In particular, for SXS:0305, ${A}_1 \sim 1$ for $t_0=0$, while ${A}_1 \sim 4$ for $t_0=25 M$. 

%%%%%%%%%

%%%%%%%%%%%%%%%%%%%%%%%%%%%%%%%%%%%%%%%%%%%%%%%%%%%%%%%%%
\section{Determining the RD start time for BH spectroscopy}\label{sec:time}
%%%%%%%%%%%%%%%%%%%%%%%%%%%%%%%%%%%%%%%%%%%%%%%%%%%%%%%%%
Whether one interprets the damped sinusoids used to fit the RD as physical BH QNMs or merely as basis functions that reproduce the NR-RD, one can test the no-hair theorem by performing a Bayesian PE on all the RD-model parameters --~namely $\{{A}_{i}, f_{i}, \tau_{i}, \phi_{i} \}$, where $i=1,2$ is a label for different modes/overtones~-- of an observed GW event. Such test of the no-hair theorem requires - a) estimation of frequencies and damping times that is consistent with GR predictions, and b) that the observed frequency spectrum is resolvable. A crucial issue in performing this test is the choice of the start time of the RD analysis after the peak of the GW strain amplitude.

At sufficiently late times after the peak of the waveform, one expects the $2$-tone RD model to accurately describe the NR-RD. However, there is a trade-off: on the one hand, performing BH spectroscopy at late times in the RD leads to large uncertainties in the estimation of frequencies due to the lack of SNR; on the other hand, starting the spectroscopic analysis close to the peak amplitude of the waveform leads to a biased estimate of QNM frequencies due to the inaccuracies of the $2$-tone RD model to describe the early RD, e.g. due to missing overtones and possible nonlinear effects in the model.
In Sec.~\ref{sec:statistical-err} we estimate the expected uncertainty in the recovery of the QNM frequencies as a function of SNR using a Fisher matrix framework. Subsequently, in Sec.~\ref{sec:modeling-systematics} we study the effect of modelling inaccuracies on the recovery of QNM frequencies. 

We argue that an optimal time to start a spectroscopic analysis of a BBH RD is the time (after the peak amplitude strain) at which the expected bias in the recovery of QNM frequencies due to modelling error is equal to (or less than) the statistical uncertainty in its recovery. We illustrate the interplay of these two factors in Sec.~\ref{sec:determinetime} and provide a computationally inexpensive prescription to find the optimal start time for spectroscopic analysis of RD in Sec.~\ref{sec:prescription}. Again we study this problem quantitatively for two representative systems~\cite{Boyle:2019kee}: i) SXS:0305, and ii) SXS:1220.  

\subsection{On the statistical error}
\label{sec:statistical-err}

The SNR $\rho$ of an observed event dictates the uncertainty in the PE due to the detector noise; we refer to this uncertainty as the statistical error in the estimation of the parameter. 

\begin{figure}[h!]
\includegraphics[width=0.45\textwidth]{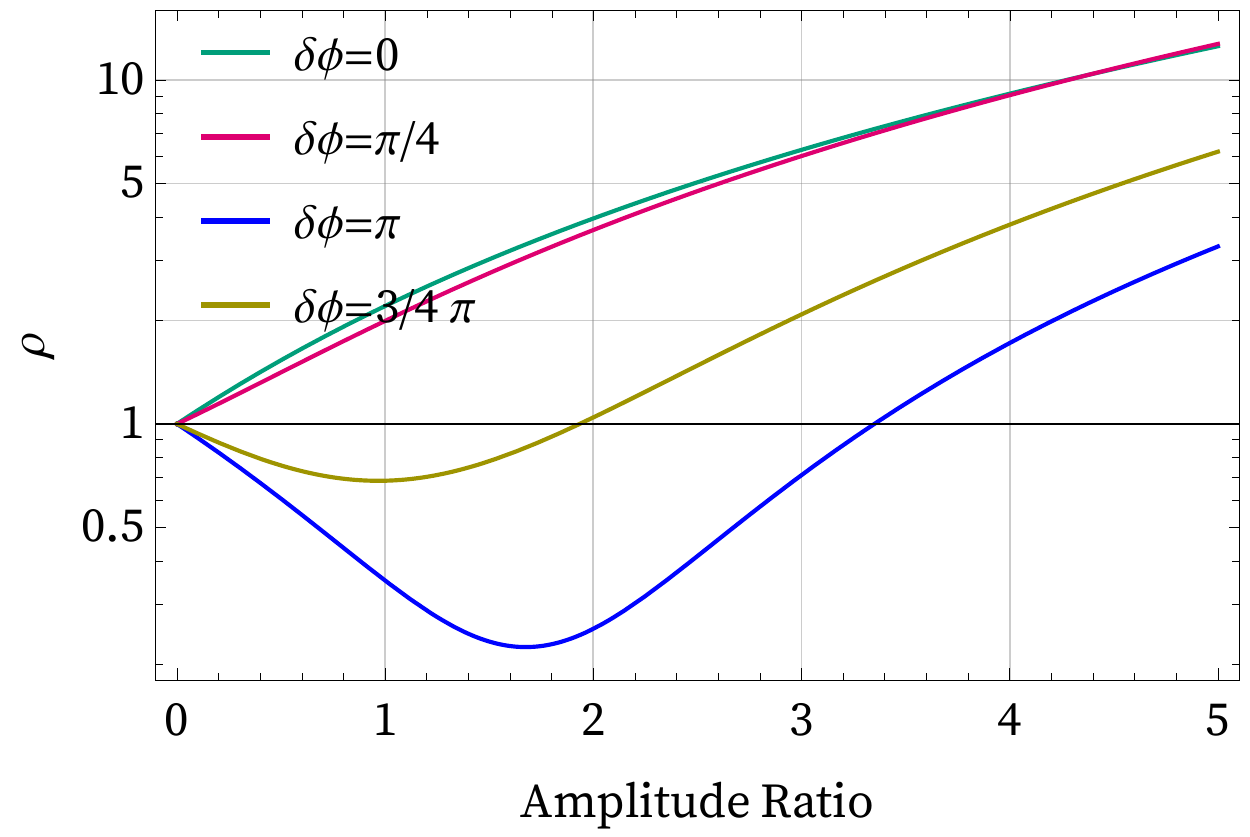}
\caption{SNR as a function of the amplitude ratio ${A}_{1}/{A}_{0}$ for different phase differences. The SNR is normalized such that the SNR in the fundamental mode is unity. The relative phase difference between the overtones has an appreciable effect on the overall RD SNR.}
\label{fig:SNR-inphase-outofphase}
\end{figure}

For a fixed RD start time, adding the overtones can markedly increase or decrease the SNR of the RD signal. This is illustrated in Fig.~\ref{fig:SNR-inphase-outofphase} where we plot the SNR (normalized to the single-mode case) as a function of the amplitude ratio between two the overtones and for various choices of the phase difference $\delta\phi$. We see that the $n=1$ overtone can notably contribute to the SNR when the amplitude ratio is $\geq 1$ and $\delta \phi \sim 0$ . 
Furthermore, the relative phase difference between the overtones plays a crucial role to the RD SNR. While a counter-phase overtone decreases the overall SNR significantly, an in-phase overtone increases the SNR by a few times (for a given amplitude ratio). This emphasises the importance of including overtones with correct relative phases for any RD analysis. 

%%%%%%%%%%%%%%%%%%%%

Next, we compute the expected statistical error $\sigma$ on the recovery of the QNM frequencies using a Fisher-information matrix framework. In a large-$\rho$ limit, the product of the SNR $\rho$ and the expected Fisher matrix spread $\sigma_{f_i}$ in the estimation of the frequency of the $i$-th QNM is a constant, i.e.,
\begin{equation}
\label{eq:kappa}
    \rho \sigma_{f_i} = \kappa_i
\end{equation}
where $\kappa_i$ depends only on the intrinsic parameters of the BBH system~\cite{Berti:2005ys}. Analytical expressions for $\rho$ and $\sigma_{f_i}$ are presented in Appendix~\ref{app:Fisher}, where we extend the previous analysis in Ref.~\cite{Berti:2005ys} to a $2$-tone RD model. 
Following Ref.~\cite{Berti:2005ys}, in the Fisher-matrix computation we assume a circularly polarized RD signal with two overtones such that i.e., ${A}_{i}^{+} = {A}_{i}^{\times}$. We set the absolute phase of the fundamental mode to zero. However, unlike in Ref.~\cite{Berti:2005ys}, we do not neglect the phases of the $n=0,1$ tones. The final analytical expression of $\sigma_{f_i}$ is cumbersome and is provided in the supplemental {\scshape Mathematica}\textsuperscript{\textregistered} notebook~\cite{webpage}.

%%%%%%%%%%%
 \begin{figure}
 \subfloat{\includegraphics[width=0.45\textwidth]{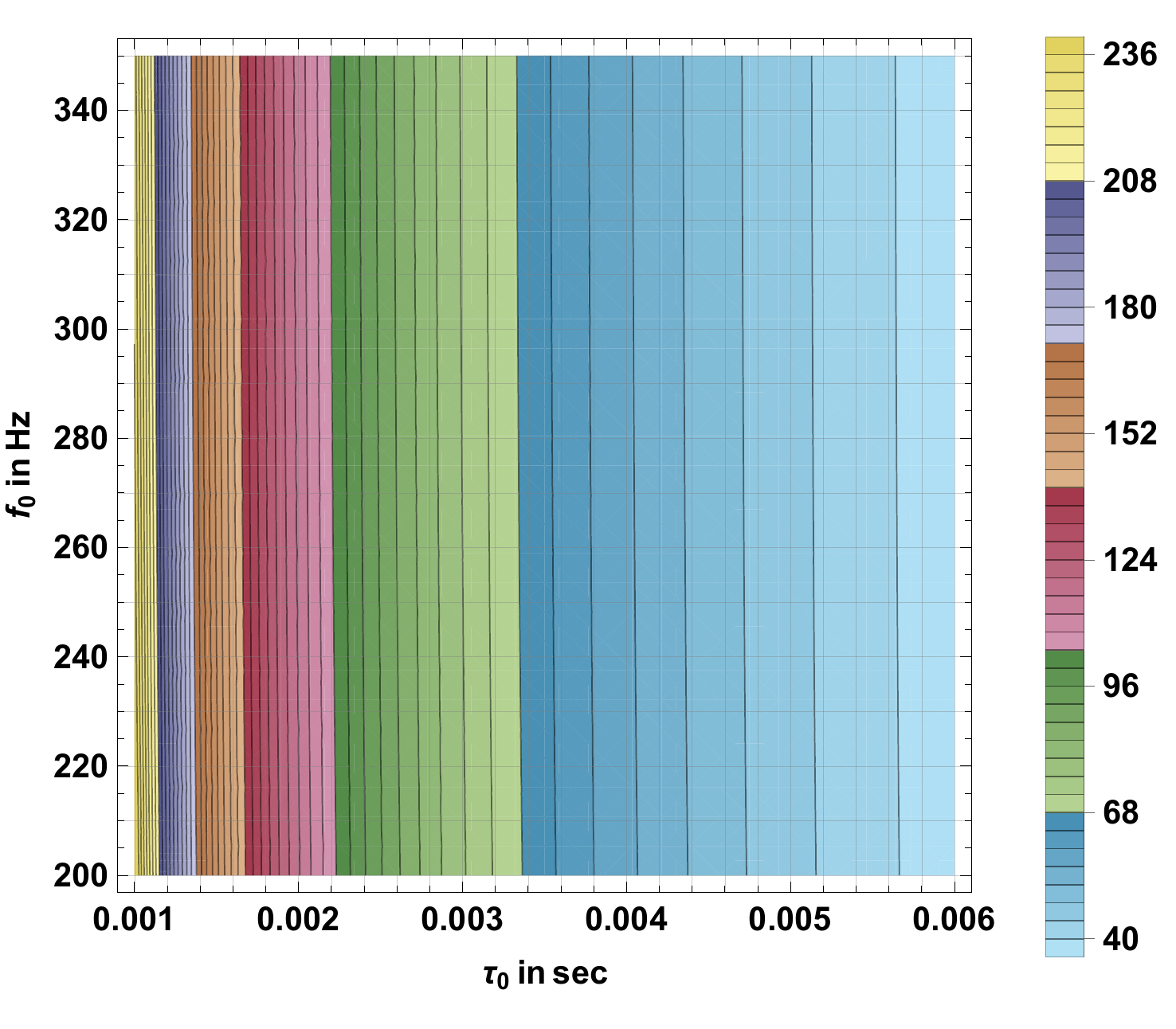}} 
 \caption{Value of $\kappa_0=\rho\sigma_{f_0}$ as function of the QNM frequency $f_{0}$ and damping time $\tau_{0}$ for $1$-tone RD model. For a given signal strength $\rho$, a larger value of $\kappa_0$ corresponds to a larger uncertainty in the estimation of  $f_{0}$. Notice that the value of $\kappa_0$ is largely determined by the damping time $\tau_{0}$ of the mode and has a much weaker dependence on the frequency. From this plot one can read the value of $\kappa_{0}$ for a given BBH system for a quick estimate of the expected uncertainty in recovering the frequency for a given RD SNR.}
 \label{fig:rho-sigma}
 \end{figure}
%%%%%%%%%%% 
 
In  Fig.~\ref{fig:rho-sigma} we study the behaviour of $\rho \sigma_{f_{0}}$ for a $1$-tone RD model as a function of the system's parameters. For simplicity we set the phase of the fundamental mode to zero, so that $\kappa_0=\kappa_0(f_{0}, \tau_{0})$ in this case. For a given value of $\rho$, we find that the systems with similar $\kappa_0$ yield similar uncertainty in the recovery of the QNM frequency. We also observe that the value of $\kappa_0$ strongly depends on the damping time but weakly depends on the frequency. 

In general, adding extra parameters to a model leads to an increase in the spread of the recovered dominant frequency  $\sigma_{f_{0}}$  due to the correlation between the parameters (see Appendix~\ref{app:correlation} for further discussions). The extent to which the uncertainty in the recovery of $f_{0}$ changes between a $1$-tone and a $2$-tone RD model depends on the intrinsic parameters of the BBH system. For simplicity, we set the phase of the dominant mode to zero\footnote{We checked that this assumption does not change the final result while simplifying the final expression substantially.}, but keep the relative phase difference $\delta \phi$ between the $n=1$ and $n=0$ tones. Therefore, in this case $\kappa_i = \kappa_i \{{A}_{1}/{A}_{0}, f_{0}, f_{1}, \tau_{0}, \tau_{1}, \delta \phi \}$.

We calculate $\kappa_{i}$ for a $2$-tone RD model calibrated to SXS:0305 and SXS:1220 using a analytical Fisher matrix calculation.  We find a strong dependence on the amplitude ratio and the relative phase difference between the two overtones. For fixed values of $\rho$ and the final mass and spin of the BH remnant, we find that $\sigma_{f_{0}}$ is maximum when the relative phase difference $\delta \phi$ is ${n \pi}/{2}$ (where $n$ is an odd integer) and minimum when $\delta \phi$ is an integer multiple of $\pi$. 

\begin{figure*}
 \subfloat{\includegraphics[width=0.47\textwidth]{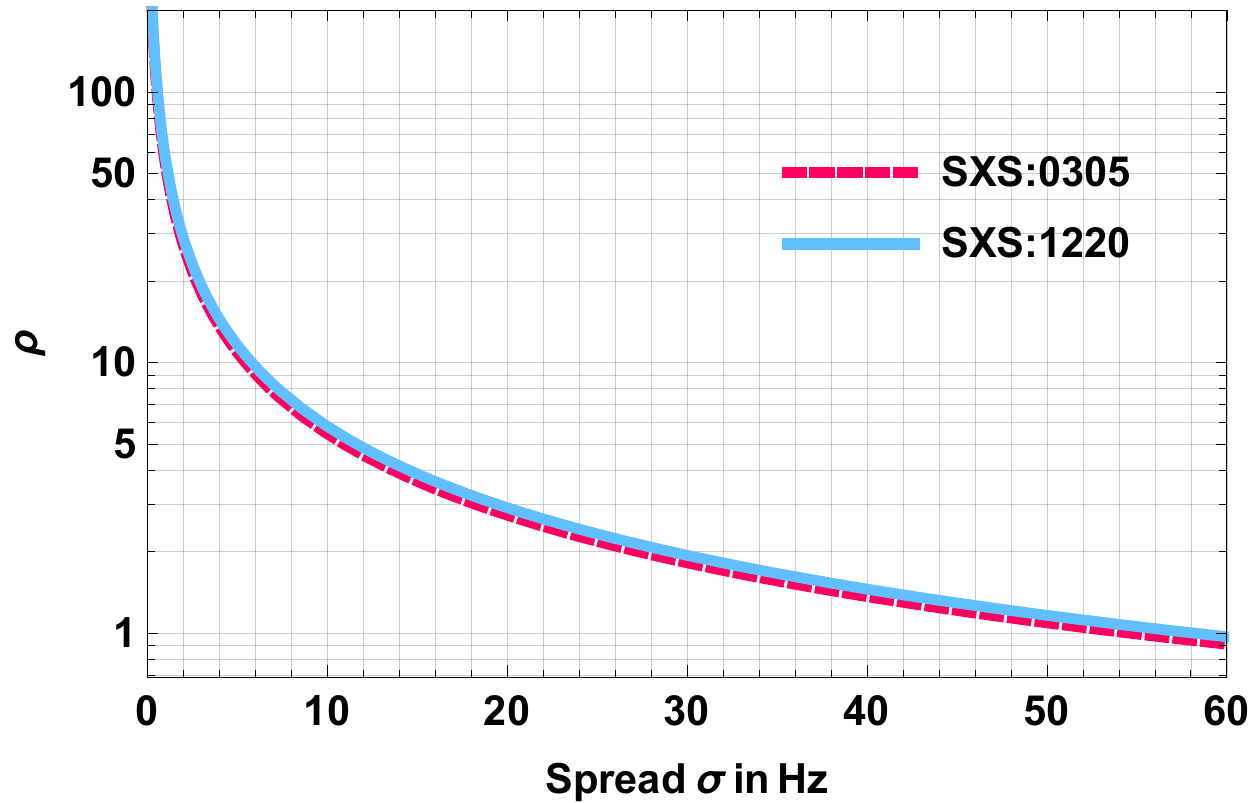}}
 \subfloat{\includegraphics[width=0.47\textwidth]{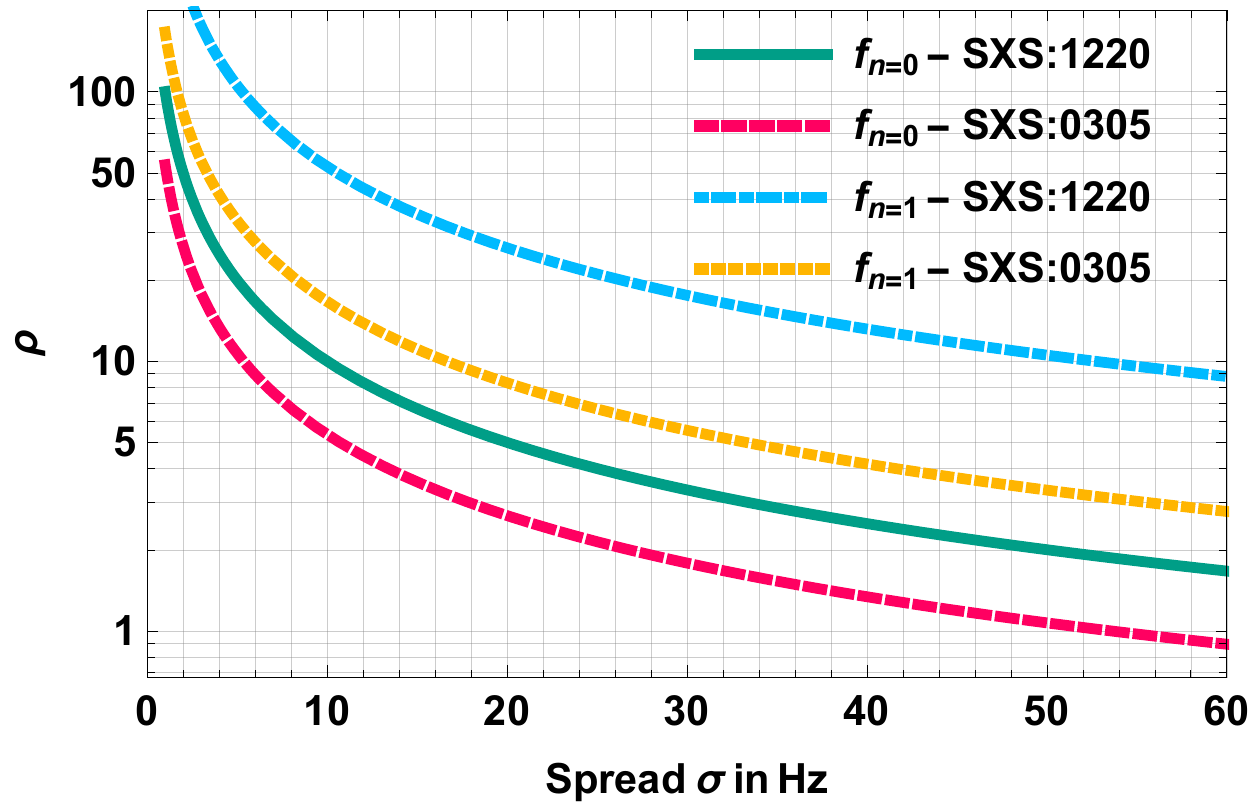}} 
  \caption{SNR $\rho$ as a function of $\sigma_{f_i}$ obtained using a Fisher matrix analysis. We show results for the $1$-tone RD model (left panel) and the $2$-tone RD model (right panel) using the best-fit values of the RD-model parameters for SXS:0305 and SXS:1220. The quantity $\kappa_i=\rho\sigma_{f_i}$ depends on the intrinsic property of the BBH system and on the parameterization of the RD model. Note that for SXS:0305 the spread in the dominant mode frequency is nearly insensitive to the addition of $n=1$ overtone to  the RD model, while for SXS:1220 the $2$-tone model produces a much larger spread in the frequency estimate when compared with the $1$-tone RD model.}
 \label{fig:rhos_Twotone}
 \end{figure*}

Using this value of $\kappa_{i}$, we study the variation of $\sigma_{f_{i}}$ as a function of RD SNR $\rho$ in Fig~\ref{fig:rhos_Twotone}. We plot the relationship between $\sigma_{f_{i}}$ of the recovered QNM frequencies and the signal strength $\rho$ for two systems, SXS:0305 and SXS:1220, scaled to $M_{f} = 68.5 \Msun$. As expected, the uncertainty $\sigma_{f_{1}}$ in the recovery of the QNM frequency of the $n=1$ overtone has a much larger uncertainty compared to $\sigma_{f_{0}}$ for a given value of $\rho$.  
Comparing the left panels ($1$-tone) to the right panels ($2$-tone) of Fig.~\ref{fig:rhos_Twotone}, we see that in SXS:0305 including the $n=1$ overtone in the RD model does not change the statistical error on the fundamental frequency significantly, unlike for SXS:1220. This is also seen in Table~\ref{tab:kappa-tab} where we present the values of $\kappa_i$ obtained for the $1$-tone and the $2$-tone RD models. The uncertainty in the frequency estimate of the fundamental mode using a $2$-tone RD model broadens and --~depending on the parameters of the BBH system~-- this may significantly worsen the estimation of $f_{0}$ \footnote{This fact is based on the estimated spread in the recovered frequency using a Fisher matrix approximation. We believe that this is due to an increased correlation between the two overtones that is dominantly affected by the frequency difference as well as the phase difference between the two tones. The dependence of this on the mass ratio will be discussed in a follow-up work soon.}. This must be considered when performing an inspiral-merger-RD consistency test which can be performed with a single QNM.

Here we highlight that $\sigma_{f_{i}}$ predicted by the Fisher matrix sets a lower bound on the estimation of the frequency $f_i$. It gives an accurate estimation of the uncertainty in the recovery only in a high SNR limit and when the noise is characterized by a Gaussian. In reality, when performing a Bayesian PE on the GW data from the current LIGO/Virgo detectors, both of these assumptions may be violated. We compare the $90 \%$ credible interval presented in Fig.~5 of \cite{TGR} for GW150914 signal to the Fisher matrix calculations using the approximate relation $\Delta_{90\%} \approx 1.64 \times \sigma_{\rm Fisher}$\footnote{The ${n-\sigma}$ spread in Gaussian noise is given by $\Delta_{p} \approx n \times \sigma_{\rm Fisher}$, where $p$ is the probability of falling inside the $n-\sigma$ 
confidence region: eg., $n=1.64$ for $p=90\%$ and $n=3$ for $p=99.7\%$.}~\cite{Ohme:2013nsa,Chatziioannou:2017tdw,Jimenez-Forteza:2018buh}. We see that the Fisher matrix underestimates the errors considerably\footnote{We also compared the Fisher matrix computation to a PE performed on a $1$-tone RD injection in zero noise. At a signal strength of $\rho = 15$, the Bayesian PE (with flat uninformative priors) estimates $\Delta_{90 \%} f_{0} \sim 8 \,\rm{Hz}$ while the Fisher matrix calculation predicts $\Delta_{90 \%} f_{0} \sim 6 \,\rm{Hz}$. Therefore, the Fisher matrix calculation agrees with the Bayesian PE in a zero-noise realisation up to $\sim 0.8 \%$ of the dominant-mode frequency.}. For instance, at a start time $t_{0} \sim 10 M$, the  $\Delta_{90 \%} $ $f_{0} \sim 18\,{\rm Hz}$~\cite{TGR}, whereas the Fisher matrix estimates $\Delta_{90 \%} $ $f_{0} \sim 10.3\,{\rm Hz}$. Therefore in the rest of the paper, we choose to consider $3 \sigma_{f_{i}}^{\rm Fisher}$ as our estimate for the uncertainty in the recovery of the frequencies to account for larger spread seen in PE.

 %%%%%%%%%
%%%%%%%%%

\begin{table}[]
\begin{tabular}{|c|c|c|c|}
\hline
         & $\kappa_{{0}}^{1-{\rm tone}}$ & $\kappa_{{0}}^{2-{\rm tone}}$ & $\kappa_{{1}}^{2-{\rm tone}}$ \\ \hline
SXS:0305 & 54.51                     & 53.53                     & 166.32                    \\ \hline
SXS:1220 & 58.09                     & 99.97                     & 526.68                    \\ \hline
\end{tabular}
\caption{The value of $\kappa_i$ for SXS:0305 and SXS:1220 calculated for a $1$-tone and a $2$-tone RD model. Note that while for SXS:0305 the value of $\kappa_0$ remains roughly constant for the $1$-tone and $2$-tone RD model, for SXS:1220 it increases nearly by a factor of $2$ in the $2$-tone RD model.}
\label{tab:kappa-tab}
\end{table}

%%%%%%%%%%%%%%%%%%%%%%%%%%%%%%%%%%%
\subsection{On modelling systematic errors}
\label{sec:modeling-systematics}
%%%%%%%%%%%%%%%%%%%%%%%%%%%%%%%%%%%
The statistical uncertainty reduces as the SNR of the signal increases. Pushing the start time of the analysis closer to the merger increase the SNR in the RD exponentially. However, describing a BBH RD close to the merger with a $2$-tone RD model introduces inaccuracies leading to a biased estimate of the QNM frequencies. In this section, we study the biases in the recovered QNM frequencies as a function of the start time for a spectroscopic analysis of RDs using a $1$-tone and a $2$-tone RD models. We use the mismatch [defined in Eq.~\eqref{mismatch}] as a measure to quantify the inaccuracies of the RD model. This is purely a mathematical measure and does not include any statistical errors due to noise in the detector. Consequently, the mismatch quantifies the bias in the recovered QNM frequencies occurring solely due to modelling inaccuracies as a function of the start time of the fit. 

We consider a n-tone RD model, allowing the frequencies of the damped sinusoids to deviate away from the QNM frequency by a quantity $\alpha_{i}$. We refer to this model as modified  n-tone RD model and is given by -
\begin{align}
 h = \Sigma_{n} {A}_{n} e^{- \iota t \omega_{n} (1 + \frac{\alpha_{n}}{100}) } e^{- t/ \tau_{n}}\,, \label{modifiedRD}
\end{align}
where $n$ is the overtone index and $\alpha_{i}$ is the relative (percent) deviation of the frequencies from the GR-QNM spectrum, $\omega_n\equiv\omega_{lmn}$.

Ideally, should the above n-tone RD model match the NR-RD perfectly, we expect $\alpha_{n} =0$, i.e., we expect a zero bias in the recovered QNM frequencies. A non-vanishing value of $\alpha_{n}$ quantifies the systematic errors of a $2$-tone model to reproduce an NR-RD.

\begin{figure}
 \subfloat{\includegraphics[width=0.45\textwidth]{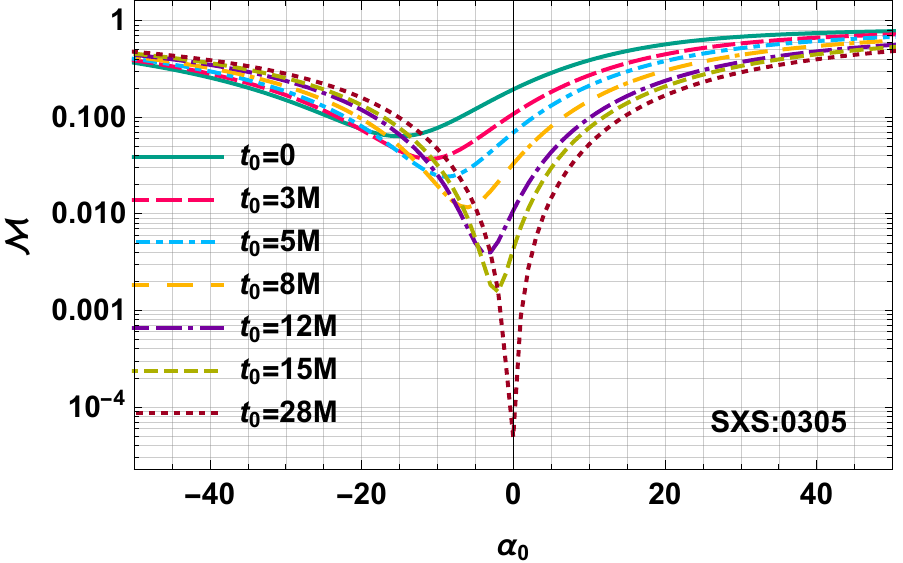}} \\
 \subfloat{\includegraphics[width=0.45\textwidth]{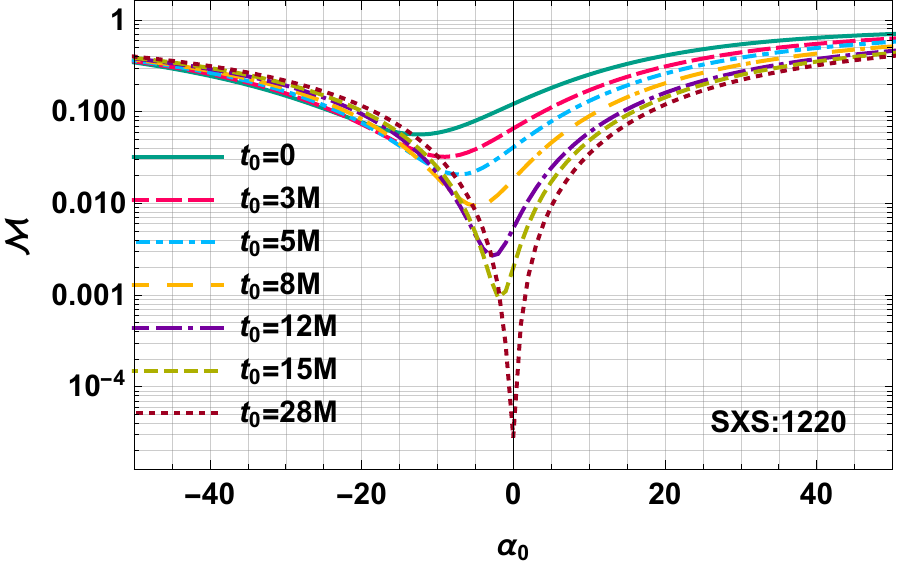}}  
 \caption{Bias incurred in a $1$-tone RD model. The top (bottom) panel corresponds to fitting the amplitude and phase of a $1$-tone RD model to SXS:0305 (SXS:1220). Different curves in the panels correspond to mismatch between the $1$-tone RD model and the NR-RD for different choices of the start time $t_{0}$. The x-axis shows the relative percentage deviation of frequency from the GR-QNM spectra. The value of $\alpha_{0}$ that corresponds to a minimum mismatch provides the estimate of bias incurred due to the inaccuracies of $1$-tone RD model to describe the NR-RD for a given start time. Note that the value of mismatch corresponding to $\alpha^{\rm{minimum}}$ is extremely small for large values of $t_0$.}
 \label{fig:alpha-match-1tone}
 \end{figure}

We first consider the $1$-tone RD model comprising of only the dominant $l=m=2$, $n=0$ mode. The issue of start time for a spectroscopic analysis using a $1$-tone RD model has been studied in the past using various techniques~\cite{Bhagwat:2017tkm,EMOP,London:2014cma,Thrane,TGR} and we re-visit this question using a mismatch analysis. We use the $1$-tone RD model given by Eq.~\ref{modifiedRD} and fit for the amplitudes and the phases of $n=0$ tone as a function of $\alpha_{0}$. In Fig.~\ref{fig:alpha-match-1tone} we show the mismatch as a function of $\alpha_{0}$ for various choices of start time $t_0$. Each point on the curves in Fig.~\ref{fig:alpha-match-1tone} corresponds to a $1$-tone RD model with the frequency of the damped sinusoid set by $\omega=\omega_{0} (1 + \frac{\alpha_{0}}{100})$. The different curves in this figure correspond to various choices of $t_0$ and the mimima in each of these curves gives a value of $\alpha_{0}$ that best models the NR-RD. Therefore, the value of $\alpha_{0}$ that minimises the mismatch quantifies the expected bias in the recovery of the frequency.  

 From Fig.~\ref{fig:alpha-match-1tone}, we note that the mismatch curves for a start time $t_0 = t_{\rm peak}$ has a bias corresponding to $\alpha_{0} = 15$ (resp., $12$) for SXS:0305 (resp., SXS:1220). We observe that as $t_0$ is pushed to a later time, the mismatch between the RD-model and NR-RD decreases, indicating that the RD model is more accurate at a later stages in the RD. As expected, we see that for a later times, both the bias and the mismatch decrease. Note the minima of the mismatch curves occur at negative values of $\alpha_{0}$ because at early times the instantaneous frequency in a BBH RD increases and at late times reaches the QNM frequency of the $n=0$ overtone.
 
\begin{figure*}
 \subfloat{\includegraphics[width=0.4\textwidth]{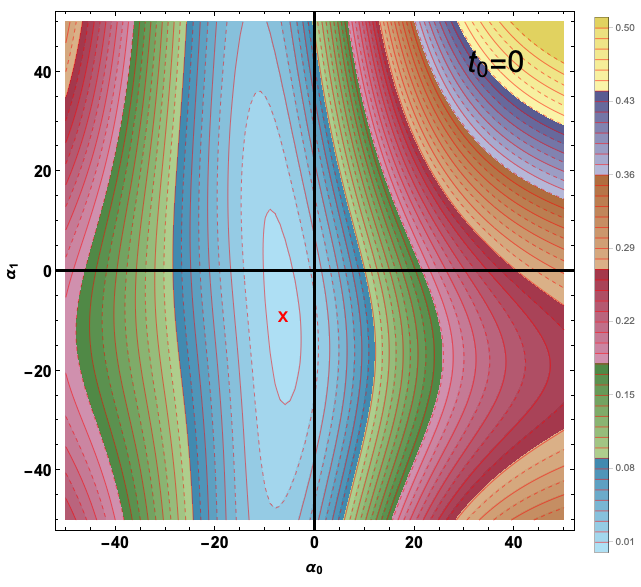}} 
 \subfloat{\includegraphics[width=0.4\textwidth]{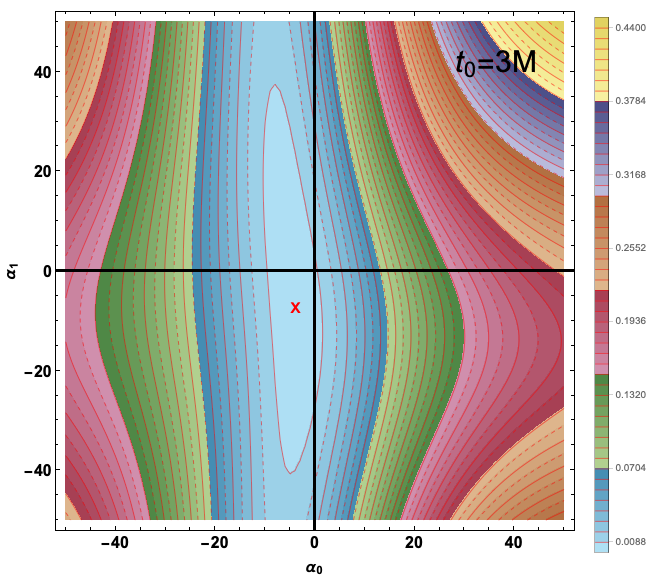}}  \\
 \subfloat{\includegraphics[width=0.4\textwidth]{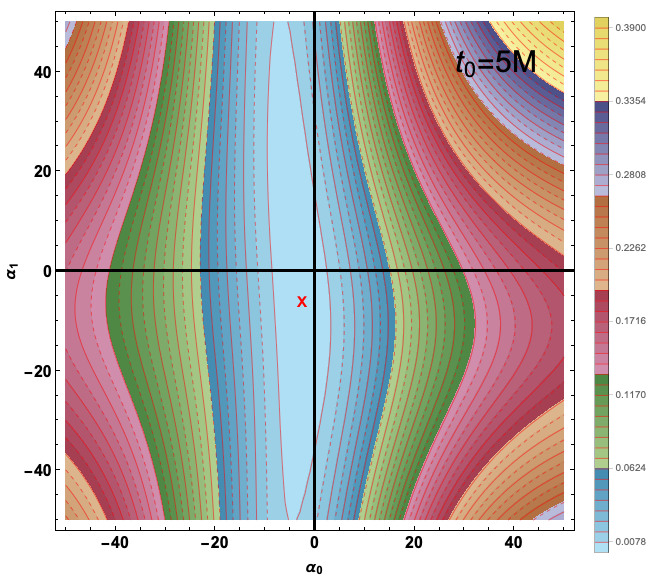}}  
 \subfloat{\includegraphics[width=0.4\textwidth]{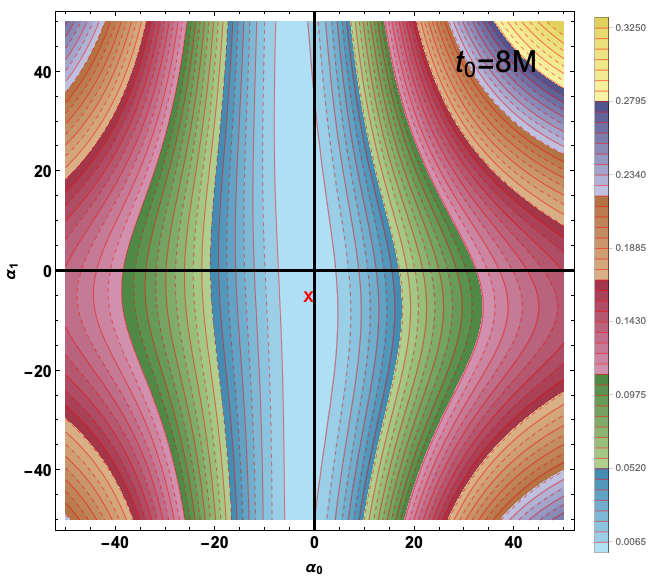}}  \\
 \subfloat{\includegraphics[width=0.4\textwidth]{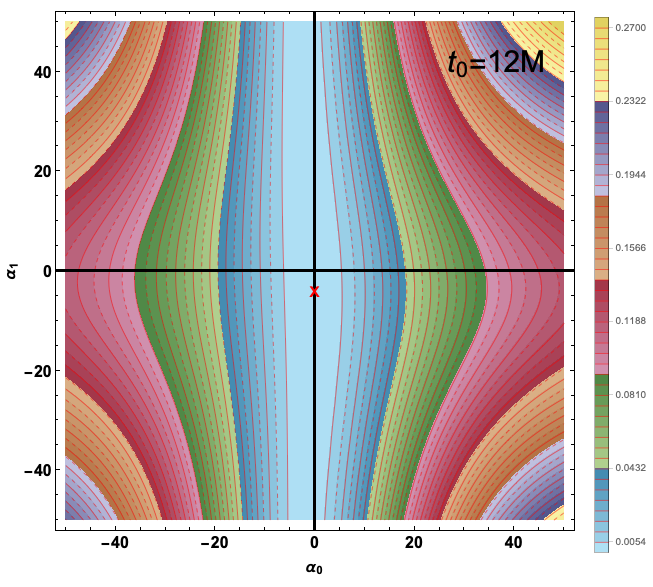}}  
 \subfloat{\includegraphics[width=0.4\textwidth]{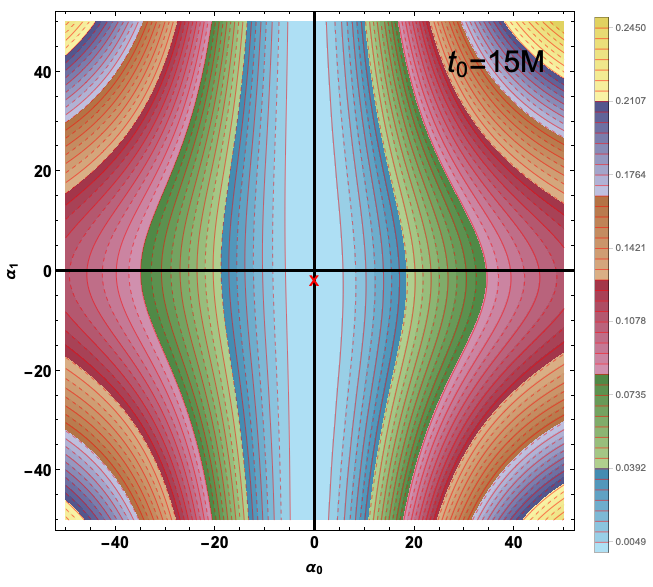}}  
 \caption{
 Bias in the frequencies for a $2$-tone RD model for SXS:0305. The frequencies of the damped sinusoids used to model the RD are allowed to vary by a relative percentage change $\alpha_{0,1}$. The mismatch between the RD model and NR-RD for each choice of $\alpha_{0,1}$ are indicated by the colour bar. The red cross marks the value of $(\alpha_{0},\alpha_{1})$ that minimizes $\mathcal{M}$ and is used to infer the bias. The six panels correspond to different starting time after the merger time. For each point in the figure, the amplitudes and the phases are obtained by performing a complex linear fit. We draw $50$ contour levels of mismatch in each panel and the innermost contour corresponds to $ {\cal M}=\{0.0109, 0.0088, 0.0078, 0.0065, 0.0054, 0.0049 \}$ for start times $t_0=\{0, 3M, 5M, 8M, 12M, 15M \}$ after the peak amplitude, respectively.}
 \label{fig:alpha0-alpha1-0305}
 \end{figure*}

\begin{figure*}
\subfloat{\includegraphics[width=0.4\textwidth]{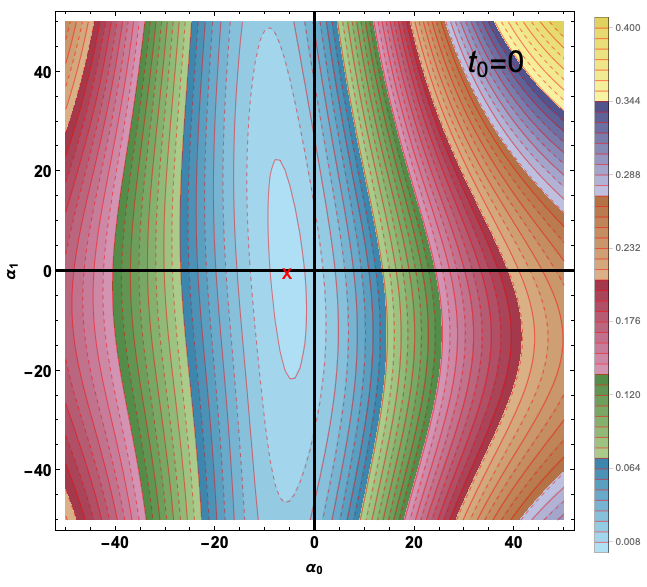}} 
 \subfloat{\includegraphics[width=0.4\textwidth]{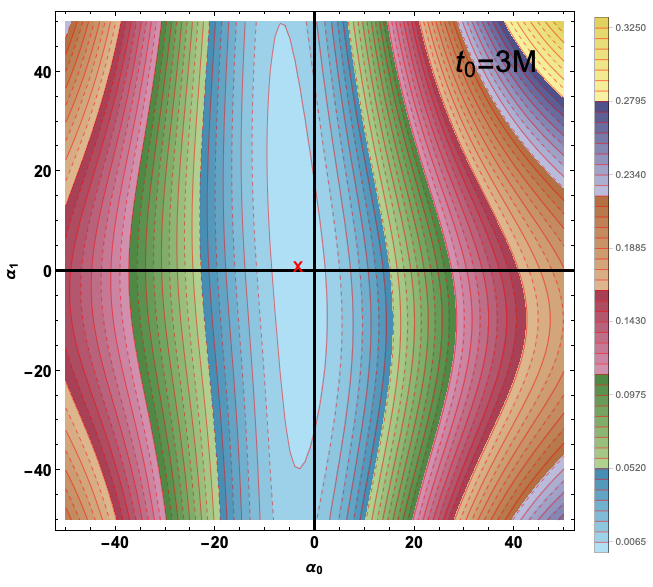}}  \\
 \subfloat{\includegraphics[width=0.4\textwidth]{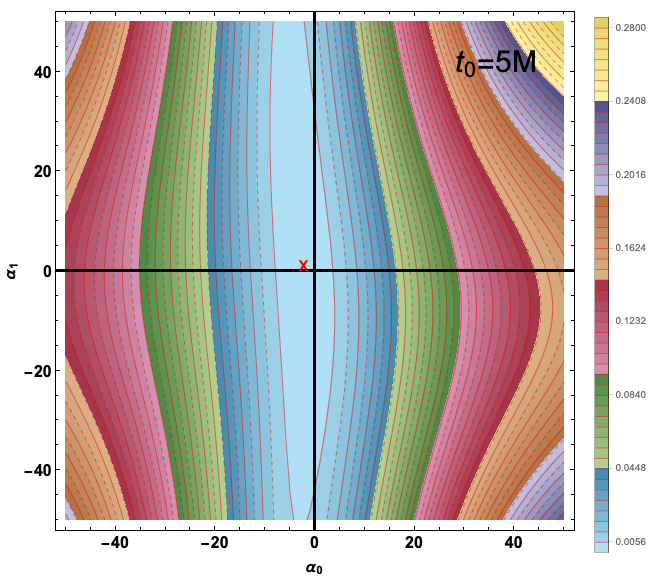}}  
 \subfloat{\includegraphics[width=0.4\textwidth]{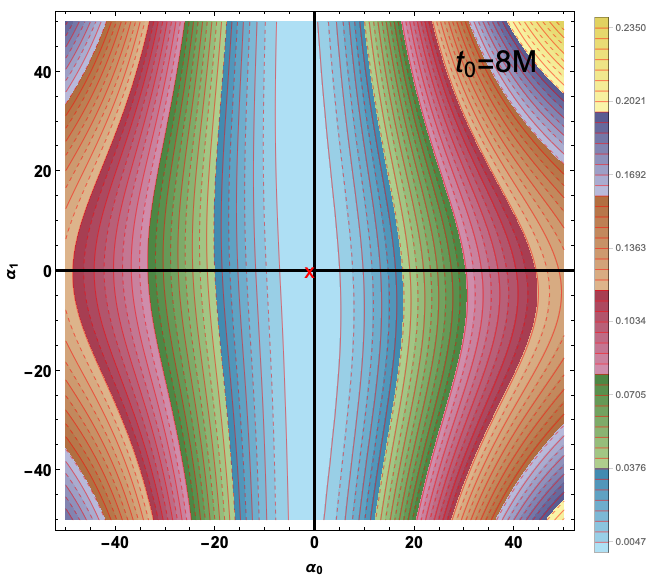}}  \\
 \subfloat{\includegraphics[width=0.4\textwidth]{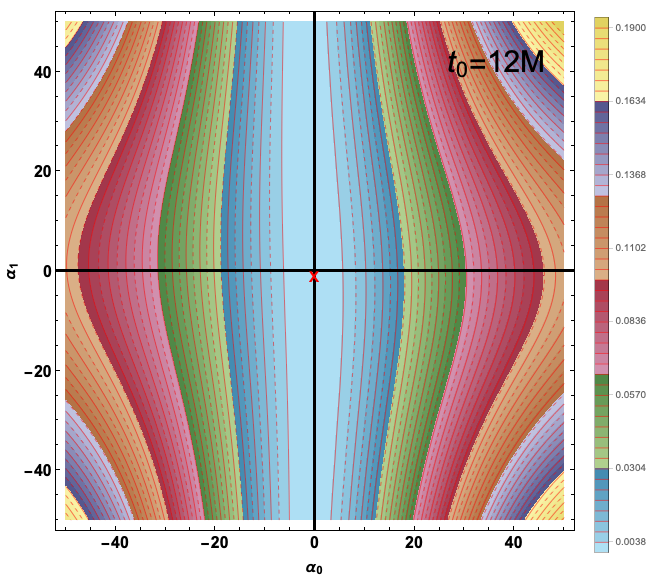}}  
 \subfloat{\includegraphics[width=0.4\textwidth]{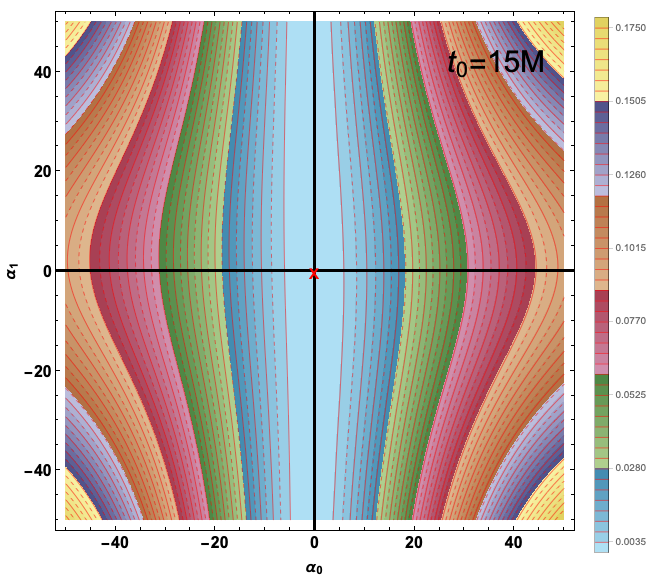}}  
 \caption{Same as Fig.~\ref{fig:alpha0-alpha1-0305} but for SXS:1220. The innermost contour corresponds ${\cal M}=\{0.0080, 0.0065, 0.0056, 0.0047, 0.0038, 0.0035 \}$ for start times $t_0=\{0, 3M, 5M, 8M, 12M, 15M \}$ after the peak amplitude, respectively.}
\label{fig:alpha0-alpha1-1220}
\end{figure*}

We conduct a similar study for a $2$-tone RD model and present its results in Figs.~\ref{fig:alpha0-alpha1-0305} and~\ref{fig:alpha0-alpha1-1220}. Figure~\ref{fig:alpha0-alpha1-0305} corresponds to fitting an NR-RD described by SXS:0305 while Fig.~\ref{fig:alpha0-alpha1-1220} corresponds to SXS:1220. Each panel corresponds to a different choice of start time and has 50 level contours of mismatch with the colour-bars indicating the value of ${\cal M}$. For each start time, the red-cross indicates the value of $\alpha_{0}$ and $\alpha_{1}$ that corresponds to the minimum mismatch. Therefore, the distance\footnote{Note that sensitivity towards the $f_{0}$ and $f_{1}$ is captured by the projection of the mismatch ambiguity contours onto the respective axes while the  bias is captured by the distance of the red cross to the axes.} of the red cross from the axis is the bias in the estimated parameters. If the $2$-tone RD model were to match the NR-RD perfectly, the red cross would coincide with the intersection of the axis (i.e., $\alpha_{0}=\alpha_{1} =0$). 

From Figs.~\ref{fig:alpha0-alpha1-0305} and~\ref{fig:alpha0-alpha1-1220} we confirm that a larger value of $t_0$ produces smaller bias. Furthermore, we see that with a latter choice of the start time (e.g., $t_0=15M$), the contours are nearly vertical, indicating insensitivity towards $\alpha_{1}$. The importance of the subdominant overtone decreases with time as a consequence of their short damping time (compared to the fundamental mode). While a large value of $t_0$ ensures an unbiased estimation of subdominant overtone frequency, one would require a large SNR to estimate the frequencies accurately from RF signal at such late start times.

Comparing the mismatch plot for the modified $1$-tone model (Fig.~\ref{fig:alpha-match-1tone}) to the contour plots for the modified $2$-tone model (Figs.~\ref{fig:alpha0-alpha1-0305} and~\ref{fig:alpha0-alpha1-1220}), we also notice that the bias in the frequency recovery of the fundamental mode is small when we allow the model to have the subdominant overtone. Interestingly, we note that this is true even when the subdominant overtone frequency itself is not well constrained (for example in SXS:1220). We find that --~from a modelling point of view\footnote{From the perspective of PE, however, it is essential to investigate the consequences of adding new degrees of freedom in the model.}~-- by allowing a degree of freedom in the form of a subdominant overtone we can decrease the bias in the recovery of the fundamental mode in the ambiguity contour plots.

Further, we note that the mismatch of the $2$-tone model is better than the fundamental mode only model, implying that the former provides a better representation of the RD signal. Also, from Figs.~\ref{fig:alpha-match-1tone},~\ref{fig:alpha0-alpha1-0305}  and~\ref{fig:alpha0-alpha1-1220} we find that for $q=4$ the $2$-tone model becomes reliable for an earlier value of $t_0$ compared to the $q =1.2$ case. This is in agreement with predictions from the perturbation theory, which is valid description even at earlier times if one of the bodies in a BBH system is much smaller than the other. 

\begin{figure}
 \subfloat{\includegraphics[width=0.45\textwidth]{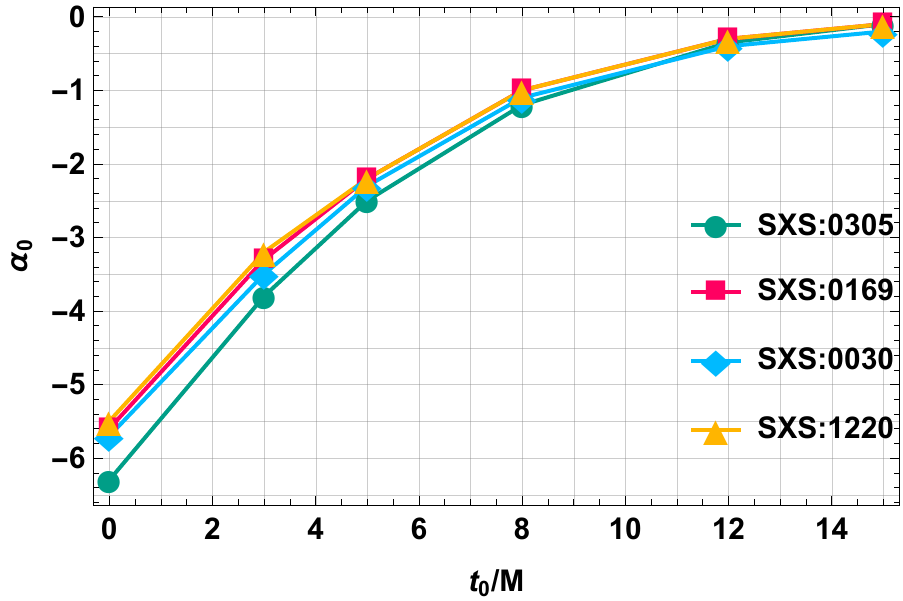}} \\
 \subfloat{\includegraphics[width=0.45\textwidth]{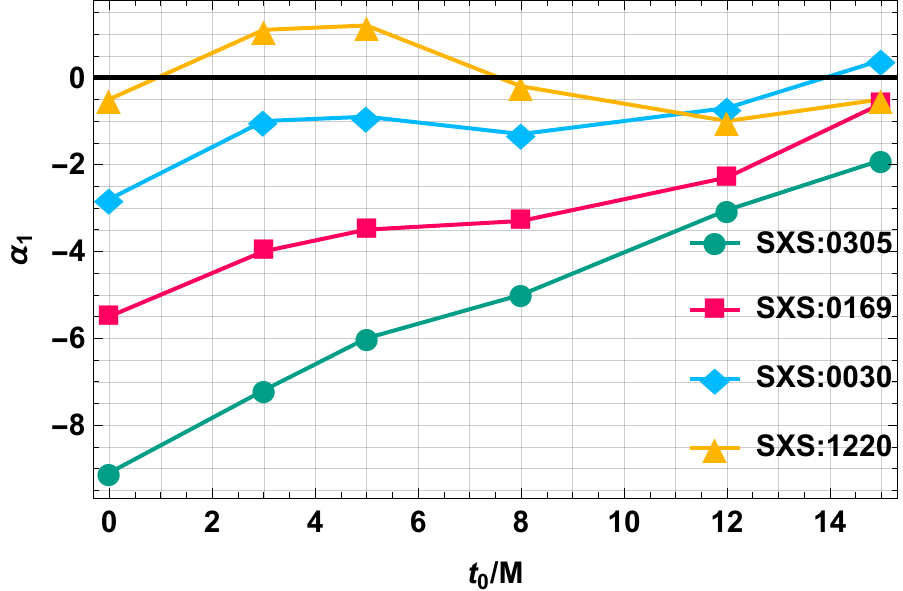}}
 \caption{The bias in $\alpha_0$ and $\alpha_1$ for different choices of starting time $t_0$. To analysis the dependence of bias in the recovery of the QNM frequencies (as a function of start time) on the mass ratio of the BBH system, in this figure we investigate 4 BBH RD  corresponding to systems with $q = \{1.2, 2, 3, 4\}$  using SXS:0305, SXS:0169, SXS:0030 and SXS:1220 respectively. We find that the bias in the recovery of the dominant mode frequency does not change noticeably with the mass ratio of the BBH system. However, the bias in the recovery of the frequency of the $n=1$ overtone frequency decreases significantly with increasing mass ratio.}
 \label{fig:bias}
 \end{figure}

We plot the bias obtained as a function of the start time for $\alpha_{0}$ and $\alpha_{1}$ in the two panels in Fig.~\ref{fig:bias}. We notice that for SXS:1220 the bias in $\alpha_{1}$ does not approach  $\alpha_{1} =0$ monotonically for late times. Therefore, to investigate this we perform this analysis for 2 more system corresponding to $q=2,3$ using SXS:0169 and SXS:0030 respectively. First, we note that the magnitude of bias $\alpha_{0}^{\rm min}(t)$ as a function of the start time  in the recovery of $n=0$ tone's frequency is similar across the 4 BBH systems with $q \sim \{1,2,3,4 \}$, However, the bias in the recovery of frequency of the $n=1$ tone shows a strong and systematic dependence on the mass ratio of the BBH system.  Larger the mass ratio, smaller is the bias obtained for a given start time of fitting. We see that for SXS:1220 the bias is small even when we start the fit from $t_{0}=t_{\rm peak}$ and fluctuation in $\alpha_{1}$ for SXS:1220 seems to arise from numerical noise. 

%%%%%%%%%%%%%%%%%
\subsection{Determining the RD start time}\label{sec:determinetime}
%%%%%%%%%%%%%%%%%

As discussed, starting the spectroscopic analysis of BH RD close to the merger leads to a biased estimation of the QNM frequencies while starting the analysis too late depletes the SNR and results in a large statistical error on the recovered QNM frequencies. We define the optimal time to start the analysis as the time at which the expected statistical spread is comparable to the bias. In this section, we use the results presented in Sec.~\ref{sec:modeling-systematics} on the expected bias and the estimate of statistical error computed in Sec.~\ref{sec:statistical-err} to provide a computationally cheap prescription for choosing an optimal start time for a BH spectroscopic analysis. We illustrate this procedure on NR-RDs corresponding to SXS:0305 and SXS:1220 waveforms. 

\begin{figure}
 \subfloat{\includegraphics[width=0.45\textwidth]{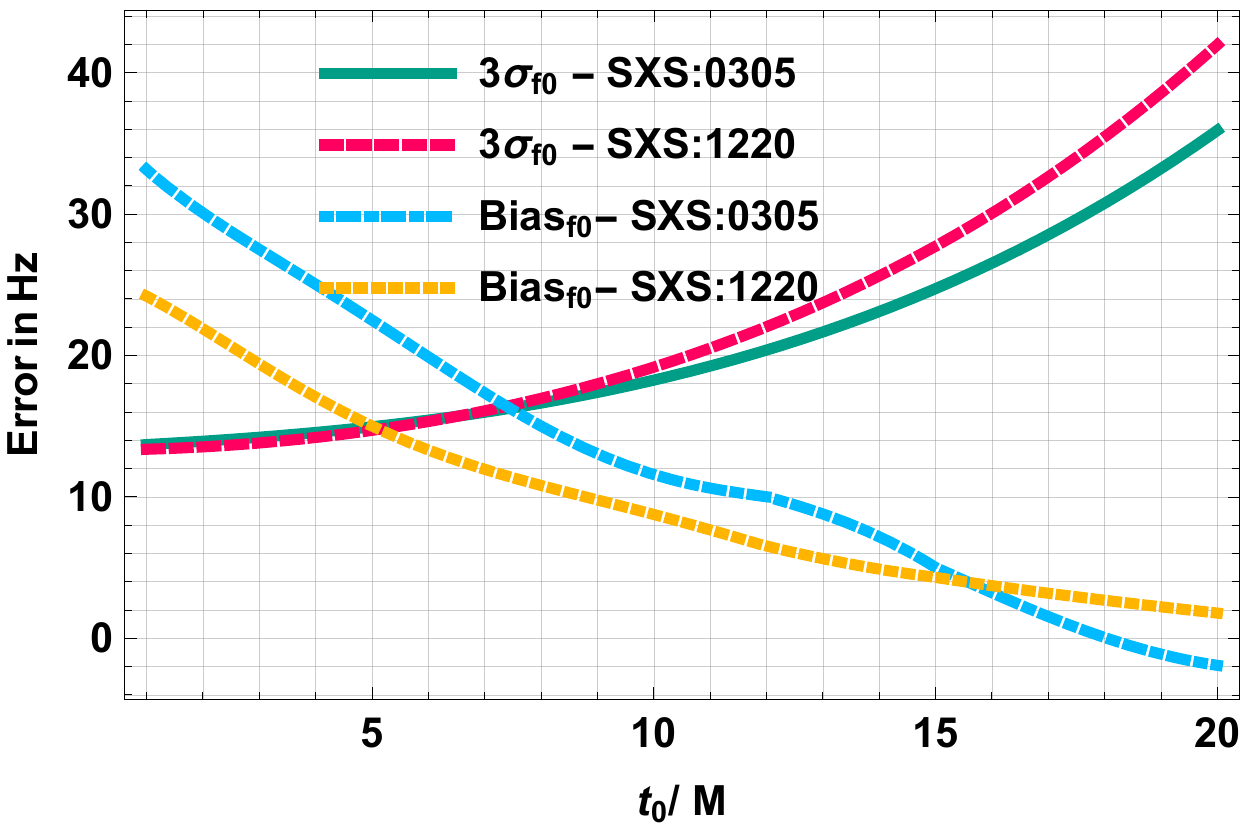}} 
 \caption{Prescription to choose the optimal RD start time for a $1$-tone RD model. The optimal start time is time at which the statistical and the systematics error become comparable. Here the two NR-RDs are rescaled to the mass of the remanent BH in the GW150914 event ($M_{f} =68.5 \Msun$). The amplitude of the NR-RD is scaled such for at $t_{0}=10 M$ the RD has an SNR $\rho\approx 8.5$. For SXS:0305  (resp. SXS:1220) we see that the optimal time to start an RD analysis is $t_{0}\sim 8 M$ (resp. $t_{0} \sim 5 M$).}
 \label{fig:starttime-one-tone}
 \end{figure}

First, we demonstrate this procedure on the $1$-tone RD model in Fig.~\ref{fig:starttime-one-tone}. We compute $\rho$ as a function of start time $t_0$ numerically and normalise it such that the SNR for a start time $t_{0}=10 M$ is 8.5. Using $\rho(t)$ in Eqs.~\ref{eq:kappa} we then estimate the expected spread in the $3\sigma$ credible interval as a function of the start time.

Next, we estimate the bias as a function of start time using Fig.~\ref{fig:alpha-match-1tone}.
% and Table~\ref{tab:alpha0-snr-tstart}. 
Finally, in Fig.~\ref{fig:starttime-one-tone} we plot the expected spread in the $90 \%$ credible interval for a RD and the expected bias as a function of the start time. The optimal time to start the spectroscopic analysis in RD is when the expected spread in the $90 \%$ credible interval is equal to the expected bias. 

To check the sensitivity of the optimal start time on the intrinsic parameters of the BBH, we apply the same procedure on SXS:1220 which corresponds to a $q=4$ BBH system. For this system, we find that the optimal time to start the analysis is as early as $5 M$ after the peak of the waveform. Therefore, the optimal start time depends on both the system's intrinsic parameters as well on the SNR contained in the BBH RD. As a general rule of thumb,  one expects that with increasing mass ratio, the post-merger relaxes to a quasi-linear configuration quicker, and therefore, the QNM description is accurate. Consequently, the optimal start time for a given SNR is closer to the waveform peak for higher mass ratio BBH systems. This picture is consistent with Fig.~\ref{fig:starttime-one-tone}. 
 
\begin{figure*}
 \subfloat{\includegraphics[width=0.45\textwidth]{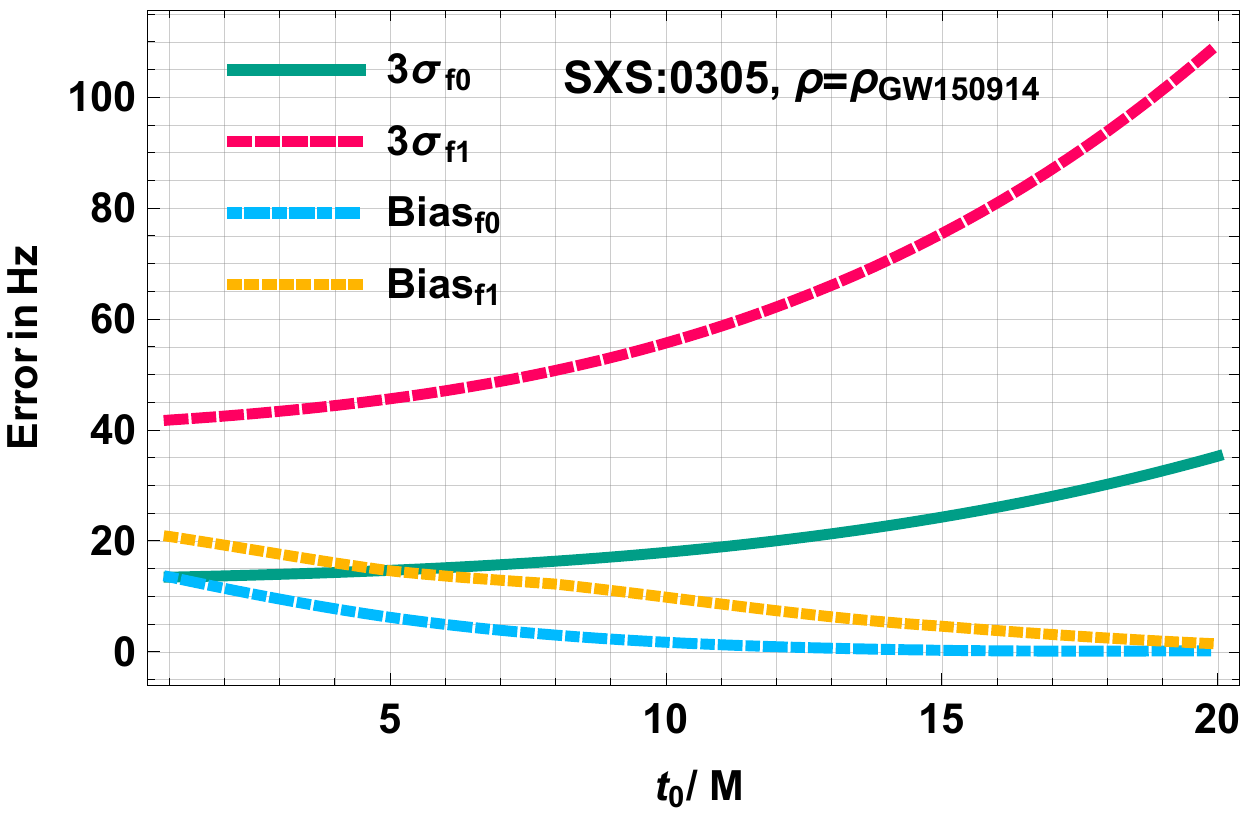}} 
 \subfloat{\includegraphics[width=0.45\textwidth]{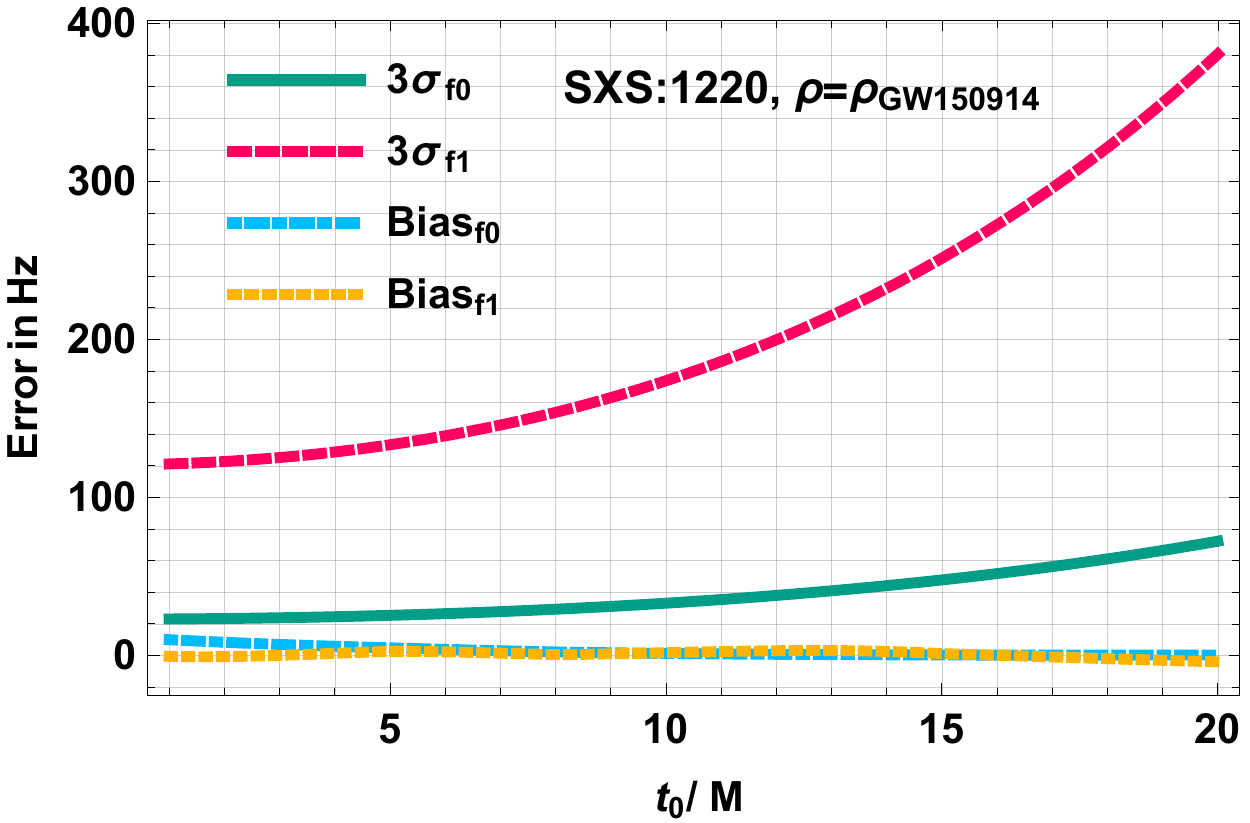}}\\
 \subfloat{\includegraphics[width=0.45\textwidth]{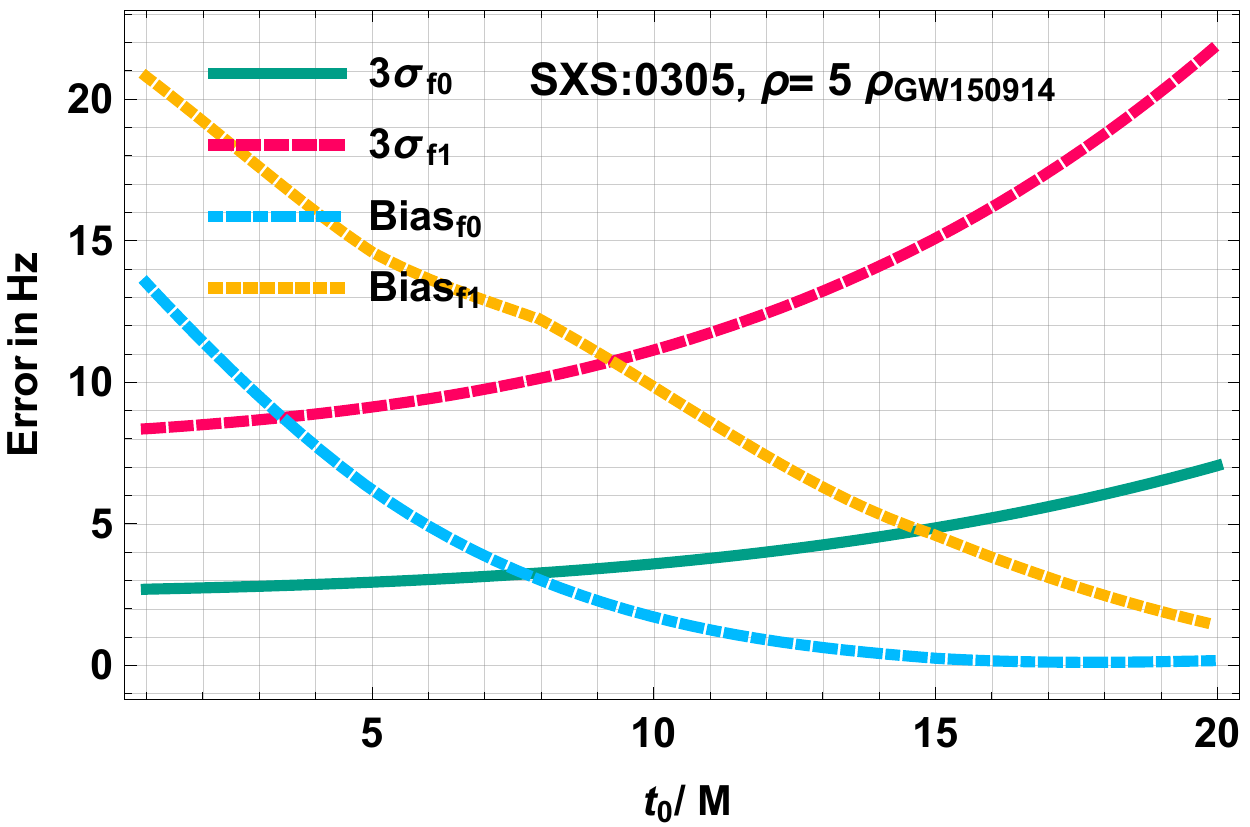}}
 \subfloat{\includegraphics[width=0.45\textwidth]{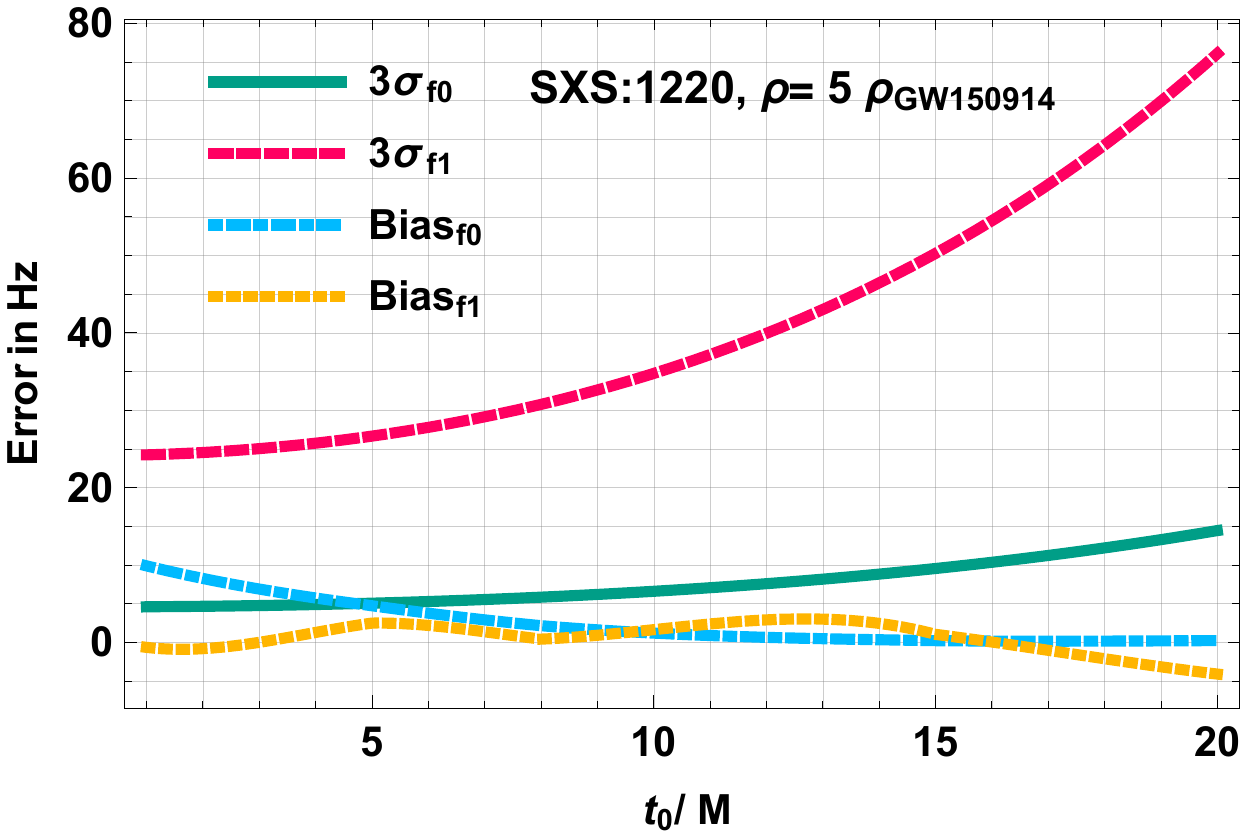}}  
 \caption{Estimation of the start time for a $2$-tone analysis. The figure shows how the start time of the analysis affects the systematic error (blue and the yellow bias curves) and the statistical error (pink and green $\Delta$ curve). The time at which the bias curves cross the $\Delta$ curve is the optimal time to start the analysis. The top left (right) panel corresponds to SXS:0305 (SXS:1220) rescaled to $M_{f} = 68.5\,\Msun$. The top  (bottom) panel correspond to the RD with an SNR of GW1501914 (5 times the SNR of GW1501914), i.e., $\rho\sim 8.5$ ($\rho\sim 42.5$) at $t_{0} = 10 M$. Note that the bias curves remains unaffected by the SNR but the statistical error curves decrease in magnitude, implying a later optimal start time. }
 \label{fig:2-ovetone-starttime}
 \end{figure*}

We implement a similar procedure for a $2$-tone RD model. The left panel of Fig.~\ref{fig:2-ovetone-starttime} corresponds to the analysis for a GW150914-like signal. Again, the optimal start time for a spectroscopic analysis is when the estimated biases in the recovered QNM frequencies of $n=0$ and $n=1$  are smaller than or equal to their estimated credible interval, i.e.,  
\begin{equation}
t^{\rm opt} = \mathrm{Max}[t_{f_{0}}^{\rm opt}, t_{f_{1}}^{\rm opt}]
\end{equation}
where $t_{f_{i}}^{\rm opt}$ is the time at which the bias in $f_{i}$ equals the spread $\Delta f_{i}$. From Fig.~\ref{fig:2-ovetone-starttime} we find that the optimal start time for performing a $2$-tone analysis is $t_0\sim 4 M$ for a GW150914-like system. 
Note that this value of $t_0$ is \emph{smaller} than in the single-mode case, showing that the inclusion of an overtone allows to start the RD analysis earlier.
In the right panels of Fig.~\ref{fig:2-ovetone-starttime}, we note that for SXS:1220 one can start the analysis at the peak of the waveform, i.e. $t_0=0$.
If the SNR in the RD increases, the expected spread in $\Delta f_{i}$ decreases while the bias is independent of the SNR; therefore, for a higher SNR RD a later optimal time is expected. From the bottom-left panel of Fig.~\ref{fig:2-ovetone-starttime}, for an equal mass BBH merger like GW150914 with $\rho=5\rho_{\rm GW150914}$ we predict that the optimal time to start a spectroscopic analysis with a $2$-tone RD model is at $t_0\sim 14 M$.

%%%%%%%%%%%%%%%%%%%%%%%%%%%%%%%%%%%%%%%%%%%%%%%%%%%%%%%%%%%
\section{Prescription for choosing the start time}\label{sec:prescription}
%%%%%%%%%%%%%%%%%%%%%%%%%%%%%%%%%%%%%%%%%%%%%%%%%%%%%%%%%%%
Here we provide a summary of the prescription for finding the optimal RD start time after the peak of the strain amplitude in order to perform a reliable spectroscopic analysis of the RD. We stress that this procedure applies to any multi-mode (including single-mode) RD analysis, both for overtones and different angular modes.

\begin{enumerate}
\item For each GW event, produce an NR/NR-calibrated inspiral-merger-RD waveform corresponding to the maximum likelihood parameter values obtained by PE on a full-inspiral-merger-RD-waveform approximant. 
\item From this reference inspiral-merger-RD signal, compute the SNR contained in the post-merger as a function of the start time, $\rho(t_{0})$. Using $\rho(t_{0})$, compute the statistical uncertainty in recovering the frequencies $\sigma_{f_{i}}(t_{0})$ through a Fisher matrix formalism (see Eq.~\ref{eq:kappa}). 
\item To assess the systematic bias incurred due to modelling inaccuracies, for different start times $t_{0}$ in the NR-RD obtained in step~1 compute mismatch ambiguity contour plots (similar to Fig.~\ref{fig:alpha0-alpha1-0305}) using a multi-tone (or multi-mode) RD model that allows for frequencies deviations $\alpha_{i}$ from the GR-QNM spectrum, as in Eq.~\ref{modifiedRD}. The point of minimum mismatch $\alpha_{i}^{\rm min}$ in the $\alpha_{0}-\alpha_{1}$ plane gives an estimate of the bias in the recovered frequencies as a function of $t_{0}$.  Thus, $\alpha_{i}^{\rm min} (t_{0})$ is the estimate of bias as a function of start time. 
\item The optimal time to start a spectroscopic analysis is the time $t_{0}$ at which the largest of the bias is equal to the spread of the corresponding frequency. 
\end{enumerate}
Therefore, the optimal time $t_{0}^{\rm opt}$ is the earliest time such that the relation
\begin{equation}
% t_{0}^{optimal} = t_{0} ~~\mathrm{s.t}~~
\sigma_{f_{i}}(t_{0}^{\rm opt}) \geq \alpha_{i}^{\rm min}(t_{0}^{\rm opt})
% |_{t_{0}=t_{0}^{optimal}}
\end{equation}
%%%
holds for all overtones in the RD model. This procedure applies to a generic RD model with any number of modes/overtones.

\section{Minimum SNR for resolvability of the overtone frequencies} 
\label{sec:resolvability}
\begin{figure}
\subfloat{\includegraphics[width=0.45\textwidth]{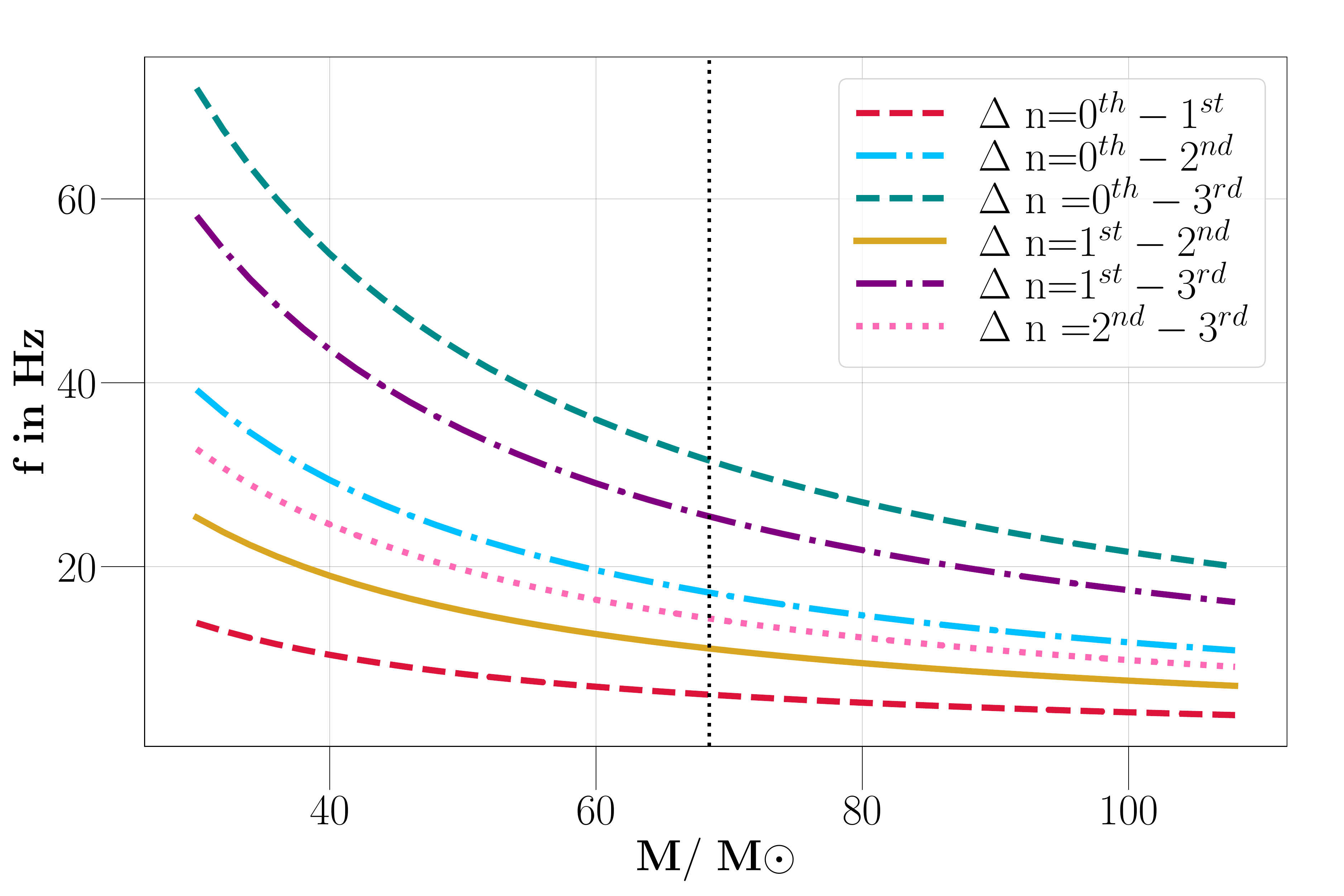}} 
\caption{Frequency difference between overtones. This figure illustrates the difference in frequencies of the first few overtones as a function of $M_{f}$ for a BH with spin $a_{f} = 0.69$. The legend $\Delta n = p -q$ denotes the difference in frequency between the $p^{\rm th}$ overtone and the $q^{\rm th}$ overtone. Note that for a GW150914-like system the frequency difference between the $n=0$ and $n=1$ overtone is only $\sim 5 \,{\rm Hz}$ and this poses challenges for resolvability. }
\label{fig:freq-diff}
\end{figure}

In this section we discuss another twist on the modelling of the RD with overtones. We assume that the $2$-tone model is accurate and investigate if we can resolve the frequencies of the two overtones apart. Resolvability is particularly challenging for overtones because their QNMs are spaced much more closely in frequency compared to the angular modes. In Fig.~\ref{fig:freq-diff} we illustrate this with an example: for a BH with mass $M_{f}= 68.5 \Msun$ and spin $a_{f} = 0.69$ the difference between the frequencies of $n=0$ and $n=1$ overtone is $\sim 5\, {\rm Hz}$, i.e.,  $f_0-f_1\leq 2\%$ of the fundamental QNM frequency\footnote{For comparison, for the same BH the $l=m=2,n=0$ QNM has a frequency $f_{220}=249.59\,{\rm Hz}$, whereas the $l=m=3,n=0$ QNM has $f_{330}=395.55\,{\rm Hz}$, see. Table~\ref{tab:l=m=2,l=m=3-freq-tau}. \label{foot:resolvability}}. The smaller the difference between the frequencies, the harder it is to resolve them. 

%[[[[For future reference: the python plots are done using Projects/higher_overtones_RD/highermode_basics.ipynb]]]]]

To address the minimum SNR required to resolve the two overtone frequencies, we use Rayleigh's resolvability criterion by demanding~\cite{Berti:2007zu}
\begin{equation}
\rho> \rho_{\rm crit} = \frac{\mathrm{Max}[\rho \sigma_{f_{0}}, \rho \sigma_{f_{1}}]}{|f_{0} - f_{1}|}\,,
\end{equation}
where $\rho\sigma_{f_i}$ is evaluated with a Fisher-matrix formalism. Note that since the overtones can have significant overlap with each other, we include the cross-terms in our calculations (unlike for the case of angular modes)~\cite{Berti:2005gp}\footnote{As discussed in Appendix~\ref{app:Fisher}, the orthogonality conditions of the spheroidal harmonics imply that the angular scalar product between two overtones with the same angular numbers is $\sim 1$. This is a crucial difference a multi-mode analysis with different angular modes~\cite{Berti:2005ys}.}. The details of the Fisher-matrix calculations are outlined in Appendix~\ref{app:Fisher}.

In Fig.~\ref{fig:Rayleigh-criterion} we show the minimum SNR $\rho_{\rm crit}$ required to resolve $n=0$ and $n=1$ overtones for a BH with mass $M_{f}= 68.5 \Msun$ for two values of the spin: $a_{f} = 0.69$ and $a_f=0.48$. The relative phase difference between $n=0$ and $n=1$ overtones is set to the best-fit value of $\delta \phi$ obtained by fitting the NR-RD of SXS:0305 and SXS:1220, respectively (using the same procedure described in the previous sections).
Regardless of the amplitude ratio between the overtones, Rayleigh's resolvability criterion estimates an SNR $\geq 30$ for a comparable-mass BH binary. Further, with the best-fit values of amplitudes and phase calibrated against SXS:0305 and SXS:1220 (scaled to $M_{f}=68.6 \Msun$), we obtain a minimum SNR $\rho_{\rm min}\sim 30$ and $\rho_{\rm min}\sim 64$, respectively.
Note that including the cross terms in the Fisher matrix calculation has a significant effect on the estimate of minimum SNR required for overtone resolvability. We find that depending on the BBH system the cross terms can either increase (as in the case of SXS:0304) or decrease (as in the case of SXS:1220) the minimum SNR by roughly a factor of $2$.

The value of $\rho_{\rm crit}$ depends crucially on the amplitude ratio and the relative phase difference $\delta \phi$ between the overtones. We illustrate this for a BH with $M_{f}=68.5 \Msun$ and $a_{f}=0.69$ in Fig.~\ref{fig:Rayleigh-criterion-contour}. We note that while the minimum SNR for resolvability is about $\rho\sim 28$ for an optimal combinations of $A_1/A_0$ and $\delta\phi$, it is as large as $\rho\sim 100$ for a large region of parameter space. When the tones have a relative phase difference $\delta \phi  = \pi$, the SNR required to resolve the overtone frequencies are $\sim$ 4-5 times smaller compared to $\delta \phi  = 0, 2 \pi$.
Further, it is worth mentioning that our results are based on a Fisher-matrix estimation of the statistical errors for a $2$-tone model. In a realistic PE we expect the errors to be larger; therefore the minimum SNR required for resolvability must be considered as an lower bound. Finally, for this study the spin of the BH is set to $a_{f}=0.69$. The relative difference in the QNM frequencies between overtones decreases for a highly spinning BH. Resolvability of overtones becomes more demanding if the final BH spins rapidly. 

\begin{figure}
\subfloat{\includegraphics[width=0.5\textwidth]{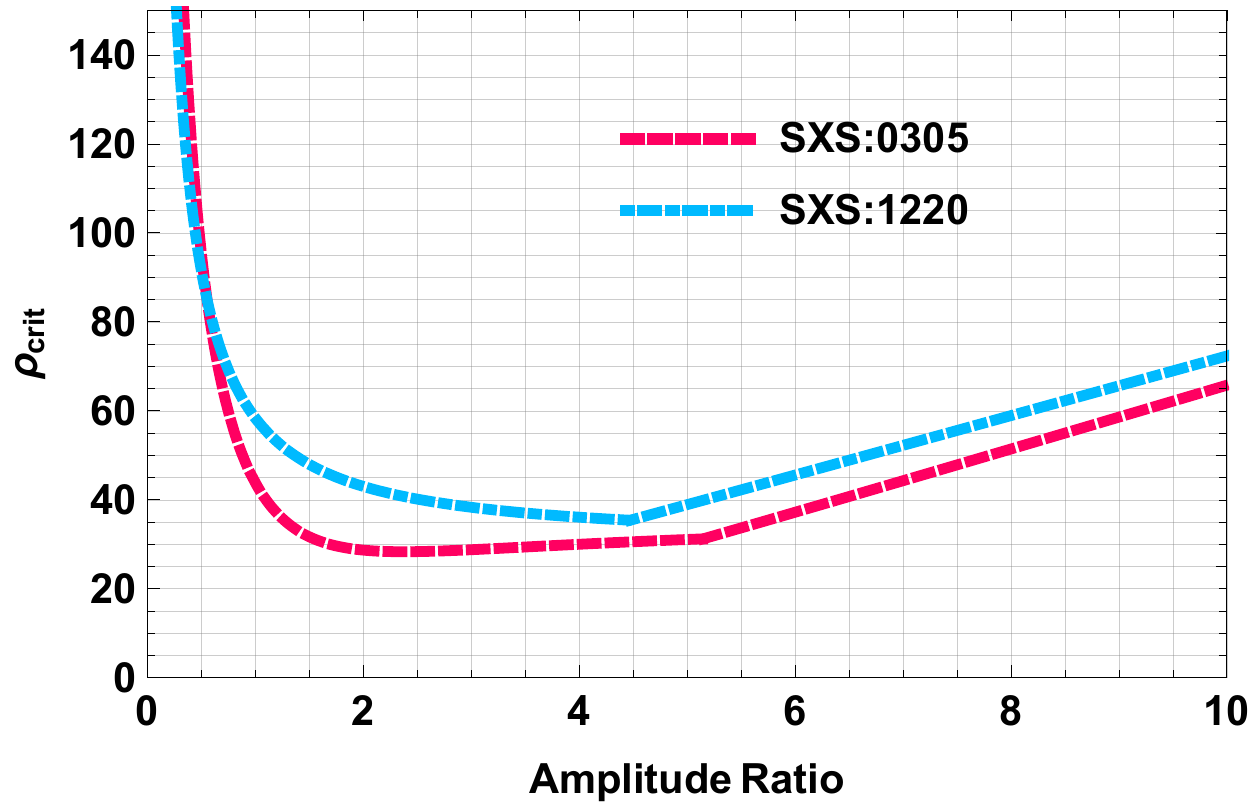}} 
\caption{Resolvability of the $n=0$ and $n=1$ overtones as a function of the amplitude ratio computed by using Rayleigh's resolvability criterion. We compute the minimum SNR required to resolve the QNM frequencies of $n=0$ and $n=1$ tones for SXS:0305 (pink curve) and SXS:1220 (blue curve) scaled to $M_{f}=68.5 \Msun$. For a GW150914-like system (SXS:0305), a minimum SNR $\rho\sim 30$ is required to resolve the $n=1$ overtone from the dominant mode. Using the best-fit values for amplitudes and phases for SXS:0305 and SXS:1220, we obtain $\rho_{\rm min}\sim 30$ and $\rho_{\rm min}\sim 64$, respectively.
}
\label{fig:Rayleigh-criterion}
\end{figure}

\begin{figure}
\subfloat{\includegraphics[width=0.45\textwidth]{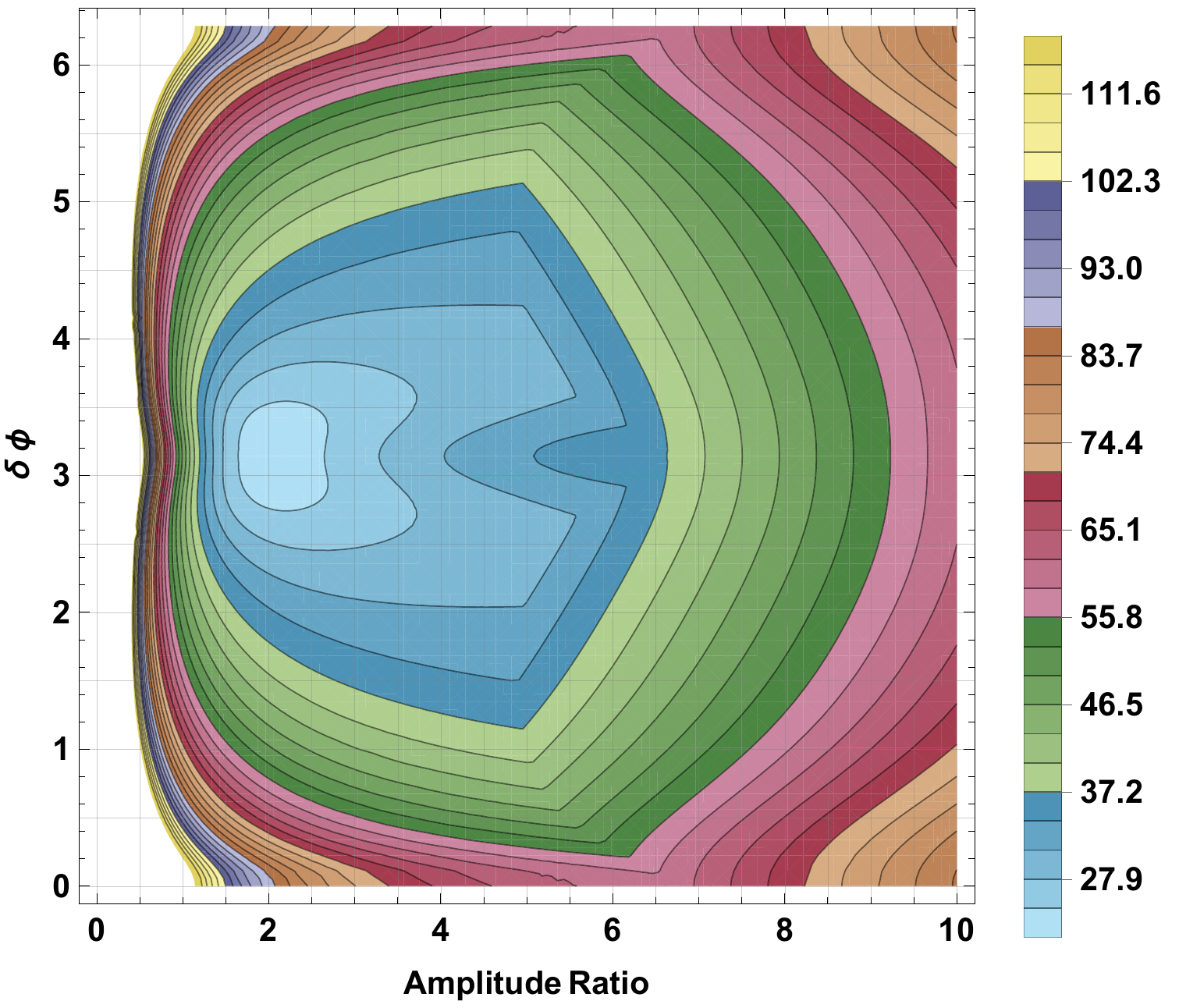}} 
\caption{Minimum SNR (obtained through Rayleigh's criterion) required to resolve the $n=0$ and $n=1$ overtone frequencies as a function of phase difference and amplitude ratio. We set the frequencies and damping times of the 2-tone RD model to the GR-QNM spectrum of a BH with $M_{f} = 0.68$ and $a_{f}=0.69$. Further, the phase of the $n=0$ tone is set to 0. Here we show the minimum SNR required to resolve the overtone frequencies as a function of the relative phase difference $\delta \phi$ and the amplitude ratio between the overtones.  We find that the minimum SNR needed to resolve the $n=1$ from the $n=0$ QNM frequency depends strongly on the relative phase difference between them. }
\label{fig:Rayleigh-criterion-contour}
\end{figure}

%%%%%%%%%%%%%%%%%%%%%%%%%%%%%%%%%%%%%%%%%%%%%%%%%%%%%%%%%%%%%%%
\section{Discussion} \label{sec:discussion}
%%%%%%%%%%%%%%%%%%%%%%%%%%%%%%%%%%%%%%%%%%%%%%%%%%%%%%%%%%%%%%%
In the previous sections we focused on investigating the technical aspects of performing the no-hair theorem tests and BH spectroscopy with overtones. 
In this section, we discuss two conceptual issues with the interpretation of frequency spectrum of the RD fits as QNMs of the BH. We conclude this section by comparing the RD analysis done with angular modes and with overtones.

%%%%%%%%%%%%%%%%%%%%%%%%%%
\subsection{On the simultaneous excitation of all the overtones}
\label{sec:start-time-simult}
\begin{figure*}
\includegraphics[width=0.95\textwidth]{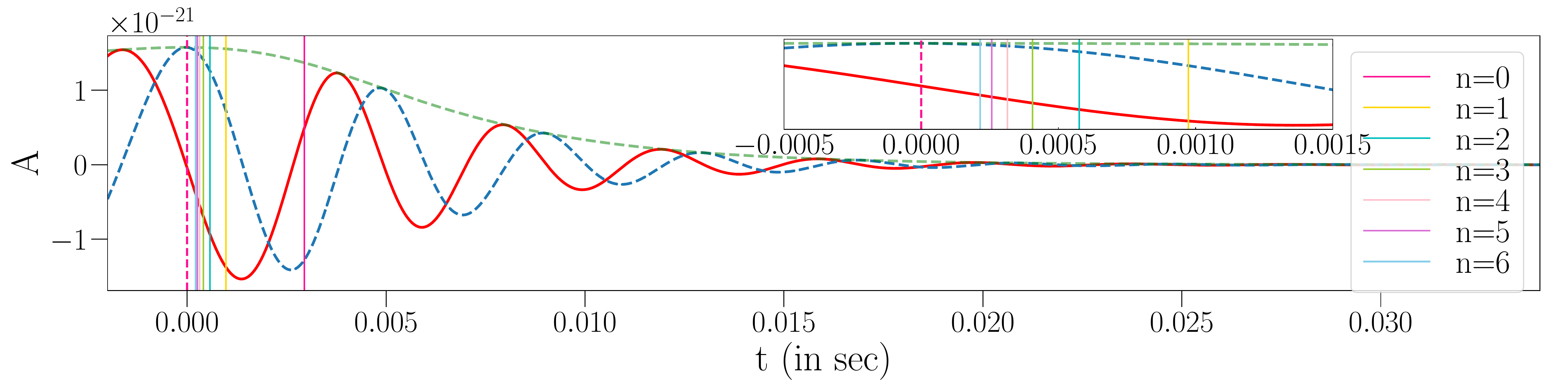}
\caption{The half-life of the first 7 tone for a GW150914-like system. The solid red curve and the dashed teal curve show the plus and the cross polarizations of the RD and the dashed green envelope is the RD amplitude. Different coloured vertical lines are drawn at 1/2-life time of the first 7 overtones assuming that all the overtones are simultaneously excited at $t_{0}=0$. Notice that the half-life of the higher overtones cluster close to the $t=0$- within the first quarter of the wave-cycle (see the zoomed-in inset).}
\label{fig:tau-on-wave}
\end{figure*}

%%%%%%%%%%%%%%%%%%%%%%%%%%
The reason why extraction of overtone frequencies was not considered conventionally to perform tests of no-hair theorem is probably due to the fact that overtones decay rapidly compared to the angular modes. The half-life of the overtones in the RD are less than half a wave-cycle as illustrated in Fig.~\ref{fig:tau-on-wave} and in Table~\ref{tab:l=m=2,l=m=3-freq-tau}. Figure~\ref{fig:tau-on-wave} shows the RD waveform (both polarizations and overall amplitude) for simulation SXS:0305 starting from the peak of the GW waveform amplitude. Assuming the RD description begins at the peak of the waveform~\cite{Matt} and assuming that all overtones start \emph{simultaneously}, the vertical lines in Fig.~\ref{fig:tau-on-wave} mark the half-life of the first few overtones (specifically, up to $n=6$). Notice that beyond $n=1$ overtone, the tones decay rapidly in a fraction of the wave cycle. 
%%%%

\begin{table}
\begin{tabular}{|c|c|c|c|c|}
\hline
$n$& $f_{22n}$ [Hz] & $\tau_{22n}$ [ms] & $f_{33n}$ [Hz] & $\tau_{33n}$ [ms]\\ \hline
0 & 249.59 & 4.1 & 395.55 & 4.0\\ \hline
1 & 244.09 & 1.3 & 392.57 & 1.3\\ \hline
2 & 233.88 & 0.8 & 386.94 & 0.8\\ \hline
3 & 220.26 & 0.6 & 379.30 & 0.6\\ \hline
4 & 205.66 & 0.4 & 370.43 & 0.4\\ \hline
5 & 197.56 & 0.4 & 361.10 & 0.3\\ \hline
6 & 197.43 & 0.3 & 351.96 & 0.3\\ \hline
7 & 198.08 & 0.3 & 343.50 & 0.3\\ \hline
\end{tabular}
\caption{List of the first overtones up to $n=7$ for $l=m=2,3$ for a reference GW150914-like system with mass $M=68.5\,M_\odot$ and $\chi=0.69$. The frequencies and the damping times quoted here are obtained using \texttt{PyCBC} package \cite{Usman:2015kfa}.}
\label{tab:l=m=2,l=m=3-freq-tau}
\end{table}

The short-lived-ness (typically contributing to less than a cycle) of overtones raise a crucial concern for modelling the RD with overtones. The models used in this paper as well as in previous work~\cite{Isi:2019aib,Matt} makes a simplifying assumption that all the overtones are excited \emph{simultaneously}. This, however, is a strong assumption and there is no first principle reasoning (for instance, from BH perturbation theory) as to why this must be the case. This assumption has been made widely in the literature in the case of modelling RD with angular modes but is better justified in that context as the angular modes live longer. Therefore, for multimode RD analysis one can pick a start time late enough such that all considered modes have all been excited. In contrast, the overtones are short-lived, posing a risk that if the overtones do not start simultaneously, a some of them can decayed to a very small amplitudes while others overtones are still being excited. This leads to several complications in the analysing and interpreting the overtone based RD-analysis. However, from a technical standpoint, allowing different modes to start at different time (naively)introduces unphysical discontinuities in the waveform or in its time derivatives.

% \pp{Maybe we can remove this paragraph.}
% An heuristic argument for the importance of allowing different modes to start at different times can be the following. For simplicity, let us assume a $2$-tone RD model and that the $n=0$ and $n=1$ modes start at $t_0^{(0)}$ and $t_0^{(1)}<t_0^{(0)}$, respectively. Then, the simplifying assumption that the modes start at the same time seems justified as long as the $n=1$ mode does not die off completely by the time the $n=0$ mode starts, i.e., whenever $ t_0^{(0)}-t_0^{(1)}\ll \tau_1$. If this is the case, choosing $t_0 = t_0^{(0)}$ does not introduce large systematic errors. We stress, however, that this argument prevents to start the RD at $t_0=0$.
% For SXS:0305, $\tau_1\sim 1.3\,{\rm ms}$, the assumption of simultaneous excitation of overtones is justified as long as a putative delay start time between modes is much smaller than ${\cal O}({\rm ms})$.
% %
% While we used the standard simplifying assumption of simultaneity of all overtones, we emphasize that this issue needs to be studied more in future works.  

%%%%%%%
\subsection{On the presence of nonlinearity in the source frame}
\label{sec:source-frame-Kerrness}
%%%%%%%
By fitting the NR waveforms with a superposition of QNM overtones, recent papers~\cite{Isi:2019aib,Matt} showed  that one can model RD arbitrarily close to the peak amplitude of the waveform. However this implies that the waveform close to the peak amplitude of waveform is fully described by the linear perturbation theory i.e., without taking non-linearities of the theory into account.  

On the other hand, the presence of non-linearities in the source frame, particularly, in a sphere of radius about $5 M$ around the final remnant BH of a BBH spacetime has been seen in previous works \cite{Bhagwat:2017tkm}. By studying the source frame evolution, it was concluded that one needs to wait for $\sim 16 M$ after the peak of the waveform to start the perturbative analysis in the asymptotic frame. Note that the analysis in \cite{Bhagwat:2017tkm} characterizes Kerr isometry in the source frame close to the BH and therefore is robust to QNM decomposition into modes and overtones as done in the asymptotic frame. While the results in \cite{Bhagwat:2017tkm} are not susceptible to the risk of overfitting and assumptions like simultaneous excitation of all overtones, it must be noted that there are error bars in mapping the gauges close to the BH to that at asymptotic infinity. 

However, these two results seem mutually inconsistent and raise the question of whether we are interpreting the results of a RD fit with overtones correctly. It is peculiar that the non-linearity observed in the source frame dynamics close to the BH does not give rise to features distinct from the predictions of the BH linear perturbation theory at asymptotic infinity. Although addressing this issue is beyond the scope of this work, we emphasize that it must be investigated carefully in a future study.

%%%%%%%%%%%%%%%%%%%%%%%
\subsection{Comparison of overtones and angular modes}
%%%%%%%%%%%%%%%%%%%%%%%%
Resolving the overtones requires a higher SNR than resolving the angular modes due to the smaller frequency separation (see Footnote~\ref{foot:resolvability}). However, for nearly-equal mass BBHs the angular modes other than the dominant $l=m=2$ QNM are weakly excited. In such cases, overtones provide an opportunity to perform the no-hair theorem tests. An interesting problem concerns identifying the BBH systems for which BH spectroscopy with angular modes is better than with overtones (or possibly identify the optimal combination between angular modes and overtones to be included in the RD template). 

We suggest that, as a rule of thumb, unless the signal is loud ($\rho\gtrsim 30$ in optimistic scenarios, higher otherwise) BH spectroscopy with angular modes provide a cleaner interpretation and should be considered the golden route. However, the feasibility of this depends on the mass ratio and the initial spins of the BBH system as they dictates the excitation factors of QNM overtones~\cite{Baibhav,Baibhav:2017jhs}.

Although the optimistic lower bound calculated using Rayleigh's criterion and a Fisher matrix formalism indicates that the minimum SNR for resolvability of two overtones ($\rho_{\rm min}\sim 30$) is marginally within the capability of LIGO/Virgo at design sensitivity, a more realistic Bayesian PE in the presence of detector noise may require a louder RD signal ($\rho\sim 100$) to perform BH spectroscopy with overtones. These values of SNR in the RD will be achievable with the future GW detectors like LISA~\cite{Audley:2017drz} and third-generation~(3G) ground-based GW interferometers (the Einstein~Telescope and the Cosmic Explorer~\citep{Hild:2010id,Evans:2016mbw,Essick:2017wyl}).

LISA is a proposed space-based GW detector sensitive to the super-massive BBH coalescence where the BBH population is expected to have a wide range of progenitors mass ratios and spins. This is to be contrasted with the ground-based GW detectors which are sensitive to stellar-origin BHs whose population is largely comprised of binary BHs of comparable mass. A large fraction of RD signals seen by LISA are expected to be generated by unequal-mass binaries; for these systems the angular modes are sufficiently excited to allow for an accurate BH spectroscopy. On the other hand, the 3G GW detectors are likely to detect nearly-equal mass BBH coalescences at a higher SNR, and therefore, the overtones can be instrumental to perform a no-hair theorem test with ground based detectors.

Finally, in the case of neutron-star-BH coalesces the mass ratio departs significantly from unity and therefore we expect that angular modes be sufficiently excited. It would be interesting to study the QNM excitation in these systems and check the role of overtones in the spectroscopy of a neutron-star-BH merger.

%%%%%%%%%
\subsection{Future work}
%%%%%%%%%
In this study we have considered non-spinning BBH systems with mass ratios $q=1.2, 4$. In the future we plan to perform a more systematic study in order to understand the role of overtones for difference mass ratio systems. Also, the initial spins of the BBH system can have a significant influence on the excitation factors as shown in Ref.~\cite{Baibhav,Baibhav:2017jhs}. 

Furthermore, to quantify biases we have considered the systematic errors on the QNM frequencies without investigating the possible systematics in the recovery of the damping times. Our prescription to choose the optimal RD start time can easily be extended to include the biases in the damping times. Adding more modes or a mixture of angular modes and overtones to the RD-model is another interesting extension of this work.
Finally, since SNR required for overtone resolvability is expected to be higher than that predicted by the Fisher matrix estimates (optimistically $\rho \sim 30$), BBH RD events that allow for a spectroscopic analysis with overtones maybe beyond the reach of the LIGO/Virgo detectors at their current sensitivity. Therefore, it would be pragmatic to explore the possibility of stacking multiple BBH events and validating the BH QNM spectrum statistically using the RD overtones across the detected GW events.

In conclusion, we reiterate that for equal mass BBH systems where the angular modes are not sufficiently excited overtones provide a lucrative prospect for performing no-hair theorem tests. However, to resolve the overtone frequencies from a single BBH observation --~and therefore performing actual BH spectroscopy~-- we estimate that a RD SNR of $ \sim 30$ is required. This is an optimistic \emph{lower} bound and it is likely that resolving two overtones in the GW data using a fully Bayesian PE requires a much higher SNR. On the other hand, since population studies indicate a large probability for equal-mass stellar-origin BBHs~\cite{LIGOScientific:2018jsj}, developing a robust overtone analysis framework is extremely valuable for the current and next-generation ground-based detectors.

%%%%%%%%%%%%%%%%%%%%%%%
 \section*{Acknowledgements}
%%%%%%%%%%%%%%%%%%%%%%%
We are grateful to Emanuele Berti and Richard Brito for useful discussion and comments on the draft, and to Gregorio Carullo, Walter Del Pozzo, Matthew Giesler, and John Veitch for interesting discussion. We acknowledge support provided under the European Union's H2020 ERC, Starting Grant agreement no.~DarkGRA--757480 and by the Amaldi Research Center funded by the MIUR program "Dipartimento di Eccellenza" (CUP: B81I18001170001).

%%%%%%%%%%%%%%%%%%%%%%%%%%%%%%%%%%%%%%%%%%%%%%%%%%%%%%%%%%%%%%%%%%%%%%%%%%%%%%%%%%%%%%%%%%%%
\newpage
\appendix

%%%%%%%%%%%%%%%%%%%%%%%%%%%%%%%%%%%%%%%%%%%%%%%%%%%%%%%%%%%%%%%%%%%%%
\section{2-mode error analysis with Fisher matrix}\label{app:Fisher}
%%%%%%%%%%%%%%%%%%%%%%%%%%%%%%%%%%%%%%%%%%%%%%%%%%%%%%%%%%%%%%%%%%%%%

In this appendix we extend some analytical results presented in Ref.~\cite{Berti:2005ys} on a multimode error analysis of the RD using the Fisher information matrix. Here we provide the main results and refer the reader to Ref.~\cite{Berti:2005ys} for details.

We write the RD waveform in the time domain as
%%%
\begin{eqnarray}
 h_{+,\times} =  \sum_{i=0}^1\left(A_{+,\times}^{(i)} e^{-t/\tau_i} \sin(\omega_i t+\phi_{+,\times}^{(i)})\mathcal{Y}_i\right)\,,
\end{eqnarray}
%%%
where $\omega_i=2\pi f_i$ is the pulsation of the $i$-th tone, $\tau_i=Q_i/(f_i \pi)$ is the corresponding damping time (with $Q_i$ being the quality factor) and $\mathcal{Y}_i$ is the spheroidal harmonic corresponding to the angular numbers of the $i$-th mode. 
The measured waveform is $h=h_+ F_++ h_\times F_\times$, where $F_{+,\times}$ is the detector's antenna pattern functions.
As in Ref.~\cite{Berti:2005ys}, we average over the pattern angles. We assume $A_+^{(i)}=A_\times^{(i)}=A_i$, $\phi_+^{(1)}=\phi_\times^{(1)}=0$\footnote{We have checked that including a generic phase for the fundamental mode does not change the final result significantly but it makes the final formulas much more involved.}, and $\phi_+^{(2)}=\phi_\times^{(2)}=\phi$.
Therefore, the templates have seven parameters --~${A}_0$, ${A}_1$, $f_0$, $f_1$, $Q_0$, $Q_1$, and $\phi$. Note that we keep $f_i$ and $Q_i$ as independent parameters in the RD model without assuming their GR relations that allow one to write them in terms of the BH mass and spin only.

If we consider two angular modes (i.e., modes with the same overtone number $n$ but different angular numbers $(l,m)$), the orthogonality conditions of the spheroidal harmonics imply that the angular scalar product between two modes is practically zero~\cite{Berti:2005ys}. However, if the two modes correspond to different overtones with the same angular numbers, their angular scalar product is almost unity. In the following calculation we consider both cases.

We define the Fisher matrix $\Gamma_{pq}=\langle \tilde h_{,p},\tilde h_{,q}\rangle$, where $\tilde h$ is the Fourier transform of the time-domain RD waveform, $ \tilde h_{,p}$ is its derivative with respect to the $p$-th parameter and the scalar product is defined as in Ref.~\cite{Berti:2005ys}. Using the standard techniques~\cite{Vallisneri:2007ev}, we compute the covariance matrix and the (normalized) expected error in the recovery of the model parameters. In particular, we focus on the combination $\rho \sigma_{f_i}$. 

For the case of angular modes, we find: 
\begin{widetext}
\begin{eqnarray}
 \rho \sigma_{f_0} &=& \frac{1}{4 \sqrt{2}}\sqrt{\frac{{f_0}^3 \left(16 {Q_0}^4+3\right)}{f_1{A_0}^2 {Q_0}^7}} \sqrt{\frac{4 
{A_0}^2 {f_1} {Q_0}^3}{4 {f_0} {Q_0}^2+{f_0}}-\frac{{A_1}^2 {Q_1} \cos (2 \phi )}{4 {Q_1}^2+1}+{A_1}^2 
{Q_1}}\,,\\
 %%%%
 \rho \sigma_{f_1} &=& \frac{{f_1} \sqrt{\frac{128 {Q_1}^8+160 {Q_1}^6+40 {Q_1}^4+6 {Q_1}^2-\left(96 {Q_1}^6+16 
{Q_1}^4-6 {Q_1}^2+1\right) \cos (2 \phi )+1}{{A_1}^2 \left(-4 \left(24 {Q_1}^4+10 {Q_1}^2+1\right) \cos (2 \phi )+8 
\left(16 {Q_1}^4+20 {Q_1}^2+5\right) {Q_1}^2+\cos (4 \phi )+3\right)}} \sqrt{\frac{4 {A_0}^2 {f_1} {Q_0}^3}{4 {f_0} 
{Q_0}^2+{f_0}}-\frac{{A_1}^2 {Q_1} \cos (2 \phi )}{4 {Q_1}^2+1}+{A_1}^2 {Q_1}}}{\sqrt{2} {Q_1}^{5/2}}\,.
\end{eqnarray}
\end{widetext}

The above result extends that presented in Eqs.~(7.6) of Ref.~\cite{Berti:2005ys} to the case of a phase difference $\phi$ between the two modes.

Next, in the case of two overtones of the same angular mode, the additional cross-terms in the computation of $\rho$ and $\sigma_f$ are non-negligible. The final result for two different overtones is cumbersome and is provided in a supplemental {\scshape Mathematica}\textsuperscript{\textregistered} notebook~\cite{webpage}.

%%%%%%%%%%%%%%%%%%%%%%%%%%%%%%%%%%%%%%%%%%%%%%%%%%%%%%%%
\section{Correlations on a multi-tone RD model} \label{app:correlation}
\begin{figure}
\subfloat{\includegraphics[width=0.45\textwidth]{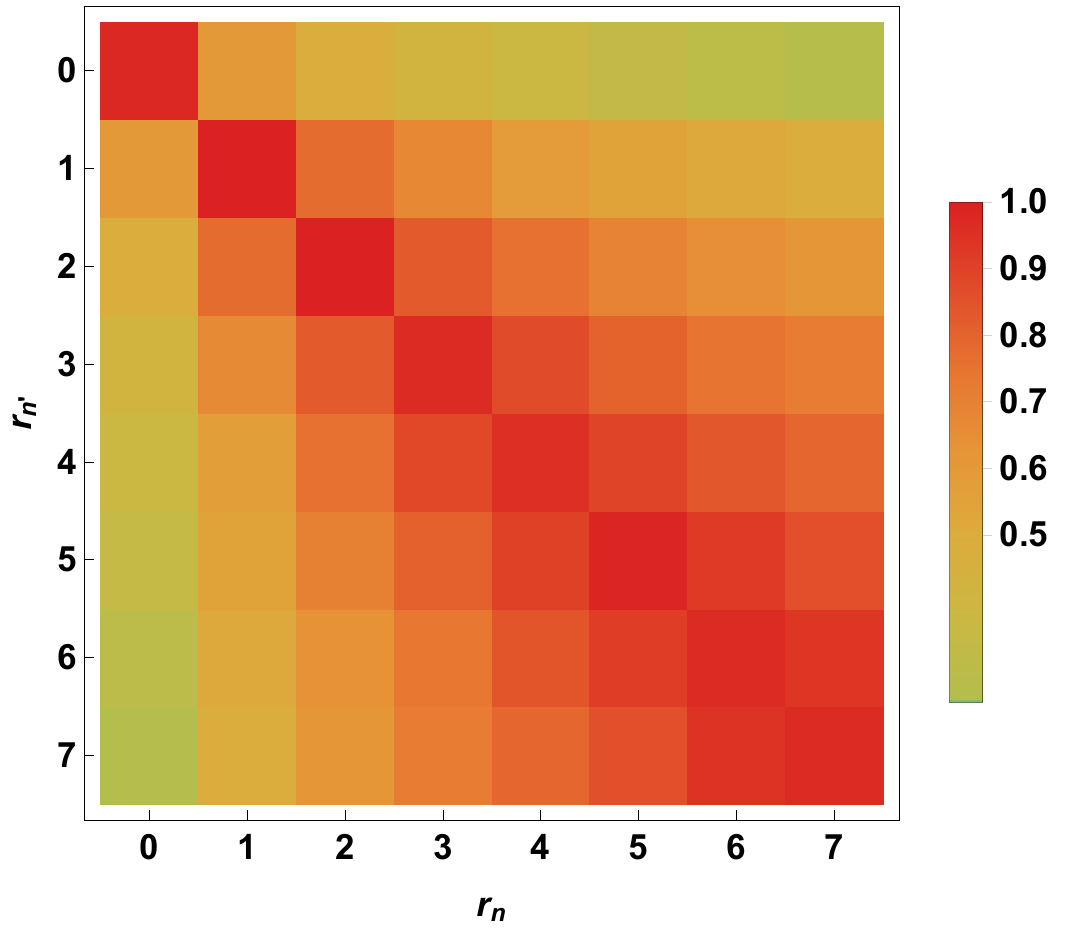}} 
\caption{Correlation matrix $corr(A_n,A_{n'})=r_{nn'}$ obtained from the 8-tone RD model used in Fig.~\ref{fig:matchplot} fitted to SXS:0305 and with start time $t_0=0$. The colour intensity of the cells increases as the correlation factors $r_{nn'}$ approach to the maximum value $r_{nn}=1$. Notice that the values of $r_{nn'}$ with $n\geq 2$ are larger as we increment the value of the tone index $n$. Disregarding the autocorrelation cells $n=n'$, we find the largest correlations occur for $r_{n n \pm 1}$. }
\label{fig:correlation}
\end{figure}

%%%%%%%%%%%%%%%%%%%%%%%%%%%%%%%%%%%%%%%%%%%%%%%%%%%%%%%%
The amplitudes of the overtones with high $n$ decay in less than one GW cycle and their short-lived-ness makes it challenging to perform robust fits to NR-RD (on the complex amplitudes of a multi-tone RD model). In Fig.~\ref{fig:correlation} we present the correlation matrix $corr(A_n,A_{n'})=r_{nn'}$ for the amplitude $A_{n}$ of the 8-tone RD model used in this study. The correlation matrix corresponds to the complex fits performed from $t=t_0=0$.

%%%%%
%%%%

Note that the correlation factors $r_{nn'}$ of the amplitudes of overtones with $n\geq 2$ are increasingly larger. We observe that higher the $n$, larger the correlation with the neighbouring values of $n$. Disregarding the autocorrelation factors, for a given overtone amplitude $A_{n}$ the highest correlation occurs with its closest neighbouring $A_{n \pm 1}$, i.e., with the elements located at the cells $n \pm 1$ out of the diagonal. In particular, the line joining the cells $r_{nn\pm 1}$ increases its intensity from $r_{32}=r_{23}\approx 0.97$ to $r_{67}=r_{76}\approx 0.99$. We note a strong correlation between the amplitudes of higher overtones. This hints towards instability of the best fit amplitude values to addition of an overtone -- casting doubt on whether the damped sinusoids used to model the NR RD be interpreted as QNM of a BH. 

\begin{figure}[!htbp]
%\subfloat{\includegraphics[width=0.45\textwidth]{Amplitude_SXS0305Random.pdf}} 
\subfloat{\includegraphics[width=0.45\textwidth]{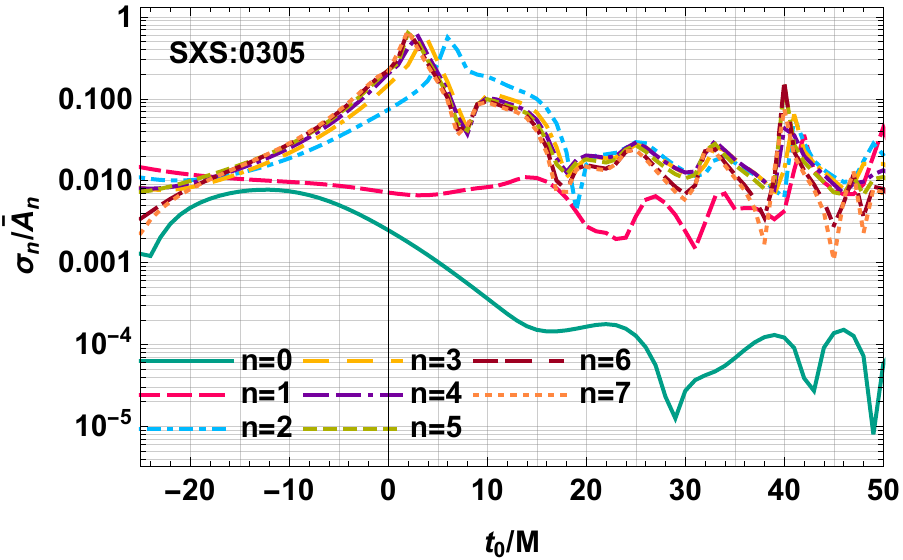}} 
\caption{Dispersion ratio $\sigma_n/\bar{A}_n$ obtained from the 100 realisations of the 8-tone RD model as a function of the starting time $t_0$. Notice that this ratio is below $1\%$ for the $n=0,1$ tones while it reaches about $75\%$ for $n>1$ and times near $t_0\sim 0$.}
\label{fig:amps_variation}
\end{figure}

In Sect.~\ref{sec:fit} we studied the effects on the mismatch produced by varying by $2\%$ the standard GR-QNM spectrum on $100$ realisations for the 8-tone RD model. Similarly, we want to quantify whether this artificial modification of the GR-QNM spectrum affects the value of the overtone amplitudes $A_n$. To this end, in Fig.~\ref{fig:amps_variation} we show the dispersion ratio $\sigma_n/\bar{A}_n$ as a function of the starting time $t_0$ obtained from the $100$ RD waveforms, where $\sigma_n$ and $\bar{A}_n$ are the standard deviation and the average amplitude of each tone respectively. $\sigma_n/\bar{A}_n$ provides an estimate of the variation on the amplitude of the tones $A_n$ sourced by modifying the GR-QNM spectrum. Notice that $\sigma_n/\bar{A}_n$ is lower than $1\%$ for tones with index $n=0,1$ and $t_0>0$. We do not observe a significant increasing along their time line.
On the other hand, amplitudes with $n>1$ tend to broaden significantly for the set of modified QNM frequencies, where the broadening is larger as the $n$ index increases and for start times near $t_0$. These values reach a maximum spread of about $75\%$.
This indicates that the best-fit values of the amplitude of higher overtones is sensitive to small changes in the frequency spectrum.

 \clearpage
\bibliography{Reference}
\end{document}